\documentclass[fleqn,usenatbib]{mnras}

\usepackage{newtxtext,newtxmath}

\usepackage[T1]{fontenc}
\usepackage{orcidlink}

\DeclareRobustCommand{\VAN}[3]{#2}
\let\VANthebibliography\thebibliography
\def\thebibliography{\DeclareRobustCommand{\VAN}[3]{##3}\VANthebibliography}

\usepackage{graphicx}	
\usepackage{amsmath}	
\usepackage{amssymb}	

\newcommand{\lamre}{\lambda_{\rm R_e}}
\newcommand{\agem}{Age_{\rm M}}
\newcommand{\agel}{Age_{\rm L}}
\newcommand{\mhalo}{M_{\rm h}}
\newcommand{\lmhalo}{\log(M_{\rm h}/{\rm M}_\odot)}
\newcommand{\lmhalonu}{\log(M_{\rm h})}

\newcommand{\lmstar}{\log(M_*/{\rm M}_\odot)}
\newcommand{\lmstarnu}{\log(M_*)}
\newcommand{\lsf}{\log(\Sigma_5/{\rm Mpc}^{-2})}
\newcommand{\lsfnu}{\log(\Sigma_5)}
\newcommand{\sigfive}{\Sigma_5}
\newcommand{\lssfr}{\log(sSFR/{\rm yr}^{-1})}
\newcommand{\lssfrnu}{\log(sSFR)}
\newcommand{\kms}{km\,s$^{-1}$}
\def \ppxf {{\small P}PXF}
\def \re {R_{\rm e}}
\def \sigmagas {\sigma_{\rm g}}
\def\kmsmpc {km$\,$s$^{-1}$Mpc$^{-1}$}
\def \oiii {[O{\small~III}]}

\def \nii {[N{\small~II}]}

\title[SAMI: spin, age, mass and environment]{The SAMI Galaxy Survey: galaxy spin is more strongly correlated with stellar population age than mass or environment}

\author[S. M. Croom et al.]{Scott M. Croom$^{\orcidlink{0000-0003-2880-9197}}$$^{1,2}$\thanks{scott.croom@sydney.edu.au},
Jesse van de Sande$^{\orcidlink{0000-0003-2552-0021}}$$^{1,2}$,
Sam P. Vaughan$^{1\orcidlink{0000-0003-2265-7727},2,3,4}$,
Tomas H. Rutherford$^{\orcidlink{0000-0002-2687-3577}1,2}$,
\newauthor
Claudia del P. Lagos$^{\orcidlink{0000-0003-3021-8564}2,5}$,
Stefania Barsanti$^{\orcidlink{0000-0002-9332-5386}2,6}$, 
Joss Bland-Hawthorn$^{\orcidlink{0000-0001-7516-4016}1,2}$, 
Sarah Brough$^{\orcidlink{0000-0002-9796-1363}}$$^{2,7}$,
\newauthor
Julia J. Bryant$^{\orcidlink{0000-0003-1627-9301}1,2,8}$,
Matthew Colless$^{\orcidlink{0000-0001-9552-8075}2,6}$, 
Luca Cortese$^{\orcidlink{0000-0002-7422-9823}2,5}$,
Francesco D'Eugenio$^{\orcidlink{0000-0003-2388-8172}9,10}$,
\newauthor
Amelia Fraser-McKelvie$^{\orcidlink{0000-0001-9557-5648}}$$^{2,5,11}$,
Michael Goodwin$^{\orcidlink{0000-0003-2955-2426}12}$, 
Nuria P. F. Lorente$^{\orcidlink{0000-0003-0450-4807}12}$,
Samuel N. Richards$^{\orcidlink{0000-0002-5368-0068}1}$,
\newauthor
Andrei Ristea$^{\orcidlink{0000-0003-2723-0810}}$$^{2,5}$,
Sarah M. Sweet$^{\orcidlink{0000-0002-1576-2505}}$$^{2,13}$,
Sukyoung K. Yi$^{\orcidlink{0000-0002-4556-2619}}$$^{14}$,
Tayyaba Zafar$^{\orcidlink{0000-0003-3935-7018}}$$^{2,3,15}$
\\
$^{1}$Sydney Institute for Astronomy, School of Physics, University of Sydney, NSW 2006, Australia\\
$^{2}$ASTRO3D: ARC Centre of Excellence for All-sky Astrophysics in 3D\\
$^{3}$Astronomy, Astrophysics and Astrophotonics Research Centre, Macquarie University, Sydney, NSW 2109, Australia\\
${^4}$Centre for Astrophysics and Supercomputing, School of Science, Swinburne University of Technology, Hawthorn, VIC 3122, Australia\\
${^5}$International Centre for Radio Astronomy Research, The University of Western Australia, 35 Stirling Highway, Crawley WA 6009, Australia\\
${^6}$ Research School of Astronomy and Astrophysics, The Australian National University, Canberra, ACT 2611, Australia\\
$^{7}$School of Physics, University of New South Wales, NSW 2052, Australia\\
$^{8}$Astralis-USydney, School of Physics, University of Sydney, NSW 2006, Australia\\
$^{9}$ Kavli Institute for Cosmology, University of Cambridge, Madingley Road, Cambridge, CB3 0HA, United Kingdom\\
$^{10}$ Cavendish Laboratory - Astrophysics Group, University of Cambridge, 19 JJ Tohmson Avenue, Cambridge, CB3 0HE, United Kingdom\\
$^{11}$European Southern Observatory, Karl-Schwarzschild-Strasse 2, 85748, Garching, Germany\\
$^{12}$AAO-MQ, Faculty of Science \& Engineering, Macquarie University. 105 Delhi Rd, North Ryde, NSW 2113, Australia\\
$^{13}$School of Mathematics and Physics, University of Queensland, Brisbane, QLD 4072, Australia\\
$^{14}$Department of Astronomy and Yonsei University Observatory, Yonsei University, Seoul 03722, Republic of Korea\\
$^{15}$School of Mathematical and Physical Sciences, Macquarie University, NSW 2109, Australia
}

\date{Accepted 7th Feb 2024. Received 5th Feb 2024; in original form 20th Dec 2023}

\pubyear{2024}

\begin{document}
\label{firstpage}
\pagerange{\pageref{firstpage}--\pageref{lastpage}}
\maketitle

\begin{abstract}
We use the SAMI Galaxy Survey to examine the drivers of galaxy spin, $\lamre$,  in a multi-dimensional parameter space including stellar mass, stellar population age (or specific star formation rate) and various environmental metrics (local density, halo mass, satellite vs.\ central).  Using a partial correlation analysis we consistently find that age or specific star formation rate is the primary parameter correlating with spin.   Light-weighted age and specific star formation rate are more strongly correlated with spin than mass-weighted age.  In fact, across our sample, once the relation between light-weighted age and spin is accounted for, there is no significant residual correlation between spin and mass, or spin and environment.  This result is strongly suggestive that present-day environment only indirectly influences spin, via the removal of gas and star formation quenching.  That is, environment affects age, then age affects spin.  Older galaxies then have lower spin, either due to stars being born dynamically hotter at high redshift, or due to secular heating.  Our results appear to rule out environmentally dependent dynamical heating (e.g. galaxy-galaxy interactions) being important, at least within 1 $\re$ where our kinematic measurements are made.  The picture is more complex when we only consider high-mass galaxies ($M_*\gtrsim 10^{11}$\,M$_{\odot}$).  While the age-spin relation is still strong for these high-mass galaxies, there is a residual environmental trend with central galaxies preferentially having lower spin, compared to satellites of the same age and mass.  We argue that this trend is likely due to central galaxies being a preferred location for mergers.
\end{abstract}




\section{Introduction}

The question of environmental impact on galaxy formation has challenged astronomers for decades.  The broad trends are well quantified, such as the morphology-density relation \citep[e.g.][]{1980ApJ...236..351D}, the reduction in star formation rates in high density environments \citep[e.g.][]{2002MNRAS.334..673L}, or the red fraction increasing with environmental density \citep[e.g.][]{2006MNRAS.373..469B,2006MNRAS.366....2W}.  These trends are understood as primarily being driven by the lower fraction of star-forming gas in galaxies occupying dense environments.  However, there is still progress to be made on understanding the detailed environmental transitions that take place.  Of particular interest is the connection between star formation and morphology.  While early-type galaxies are typically passive and late-type galaxies are typically star-forming, the mapping is not trivial or simply one-to-one.  The complexity is at least in part due to galaxy properties also depending on mass, and there is now a realization that there are independent mass and environmental quenching processes \citep{2010ApJ...721..193P}.

There is an obvious link between morphology and kinematics. This link is natural, given that the photometric structure of a galaxy is largely defined by the orbits of its stars.  \citet{2012ApJS..203...17R} found empirical relations between the specific angular momentum ($j_*$) of galaxies and stellar mass ($M_*$).  They demonstrated that early- and late-type galaxies follow parallel tracks in $j_*$ vs.\ $M_*$, but with the early-type galaxies at lower $j_*$ for a given $M_*$.  Further work found $j_*$ and $M_*$ are closely related to bulge fraction \citep{2014ApJ...784...26O} and bulge type \citep{2018ApJ...860...37S}, while \citet{2016MNRAS.463..170C} showed that at a given mass, $j_*$ is a smoothly varying function of Sersic index.  $j_*$ is also found to be dependent on gas fraction \citep{2022MNRAS.516.4043H}. 

The work of the ATLAS3D team took the relation between kinematics and morphology one step further, proposing a kinematic morphology-density relation \citep{2011MNRAS.416.1680C}.  They found a higher fraction of slow rotating galaxies in densest environments using local density (nth-nearest neighbour density), although given the volume of ATLAS3D the dense environments were largely limited to the Virgo cluster.  Slow rotators (SRs) in this context were defined as galaxies below a specific value of the $\lamre$ spin parameter \citep{2007MNRAS.379..401E}, while also accounting for ellipticity.  The kinematic morphology-density relation is qualitatively similar to the earlier morphology-density relation \citep{1980ApJ...236..351D}.  With the advent of larger samples using multiplexed integral field spectroscopy [e.g.\ the Sydney-AAO Multi-object Integral field spectrograph (SAMI) Galaxy Survey \citet{2012MNRAS.421..872C}; and the Mapping Nearby Galaxies at Apache Point Observatory (MaNGA) Survey \citet{2015ApJ...798....7B}], it became clearer that the dominant parameter that drove the fraction of SRs was in fact stellar mass \citep{2017ApJ...844...59B,2017MNRAS.471.1428V,2017ApJ...851L..33G}.

With the SAMI Galaxy Survey \citet{2021MNRAS.508.2307V} showed that, while mass appears to be the parameter most significantly correlated with galaxy spin, there is a second independent correlation with environment.  This is present both in the fraction of SRs and the mean spin of galaxies once SRs are removed.  Analysis using the MaNGA survey \citep{2019arXiv191005139G} also finds that the fraction of SRs does have a secondary dependence on environment once mass is controlled for.   

The obvious question to ask given the difference seen in galaxy kinematics as a function of environment is: what are the dominant physical processes that drive this difference?  The array of possible processes is well known and discussed in many papers.  Broadly these processes can be separated into those that are related to gas physics and those that are related to gravitational interactions.  Even though they generally do not directly change the kinematics of already formed stars, gas processes can influence the global stellar kinematics of galaxies in several ways, including suppressing the formation of new stars in dynamically cold disks.

Ram pressure stripping has been known as a possible process to remove gas for some time \citep{1972ApJ...176....1G}.  Ram pressure is important in high density regions such as galaxy clusters with many cases of  gas stripping now observed \citep[e.g.][]{2017ApJ...844...48P} in multiple phases \citep{2018MNRAS.476.4753J,2004AJ....127.3361K,2018MNRAS.480.2508M}.  Evidence of stripping is also seen in the outside-in quenching of star formation in clusters \citep{2004ApJ...613..866K} and groups \citep{2019MNRAS.483.2851S,2022MNRAS.516.3411W}.  Gas is preferentially removed first in the outer disc, leading to star formation quenching there.  In clusters ram pressure stripping appears relatively efficient.  Using phase space analysis \citet{2019ApJ...873...52O} find that galaxies with regions of recently quenched star formation are consistent with a population that has fallen into the central regions ($\sim0.5r_{200}$, where $r_{200}$ is an estimate of the virial radius of the cluster, containing a mass density 200 times the average background density) of a cluster within the last $\sim1$\,Gyr.  In groups the effect is less efficient, with \citet{2022MNRAS.516.3411W} finding that it may take several Gyr to fully quench group galaxies in this way.  See \citet{2021PASA...38...35C} for a recent review of gas stripping and quenching.

If there is not sufficient ram pressure to directly remove gas, then star formation can be quenched by stopping the supply of new gas \citep[e.g.][]{1980ApJ...237..692L}, so that the galaxy slowly decreases in star formation as the gas is used up.  This slow shutdown of star formation is expected to enhance the metallicity of galaxies, as no new pristine gas is accreted.  This scenario is consistent with passive galaxies having higher metallicities than star forming galaxies of the same mass \citep{2015Natur.521..192P}.  However, recent work by \citet{2022MNRAS.tmp.2242V} has shown that this difference is significantly reduced if metallicity is more fundamentally tied to gravitational potential rather than mass.

The above processes all influence the gas content of galaxies, which means that the dynamics of the already formed stellar populations within the galaxies should be largely unaffected.  For example, stellar kinematics are not directly impacted by gas removal.  Ram pressure can drive a temporary enhancement in star formation \citep[e.g.][]{2019MNRAS.487.4580R}, but these are likely cause only second order changes in the overall distribution of stars in the galaxy.

An alternative route that allows gas to impact stellar kinematics is the degree to which the gas forms a thin disc prior to star formation.  There is now good evidence that ionized gas is more turbulent in galaxies at high redshift \citep[e.g.][]{2012ApJ...758..106K,2015ApJ...799..209W,2019ApJ...880...48U}.  If the ionized gas traces the kinematics of the newly formed stars, then stars born at earlier epochs will be dynamically hotter.  Such a separation in age and kinematics is now starting to be seen in local galaxies \citep[e.g.][]{2019MNRAS.487.3776P}. 

In contrast to the gas-only processes, dynamical interactions can modify the orbits of the current stellar population as well as influence gas and star formation.  The most extreme of these are major mergers that can completely redistribute the orbits of stars and cause gas to flow towards the centre of a galaxy driving a burst of star formation.  Whether mergers lead to a lowering of galaxy spin largely depends on the gas content of the merging galaxies.  \citet{2018MNRAS.473.4956L} showed that dry mergers are much more effective at creating slowly rotating galaxies, as galaxies with high gas content can reform a disc. However, it is worth noting that high resolution simulations of wet mergers have been shown to form slow rotators in some cases \citep{2010MNRAS.406.2405B}.  Simulations particulary suggest that high mass slow rotators ($\lmstar\gtrsim11$) are formed through merging \citep{2022MNRAS.509.4372L}.  Given that the ability to merge is dependent on environment, it might be expected that merging could cause spin to be environmentally dependent, i.e.\ the kinematic morphology density relation.

Even if galaxies do not merge, repeated dynamical encounters may modify star formation and morphology.  \citet{2011MNRAS.415.1783B} use simulations to find that tidal interactions with other galaxies within a group environment can heat the stellar disc of a galaxy.  In clusters tidal interactions should be stronger \citep{1996Natur.379..613M} and are expected to dynamically heat galaxies as well as remove mass.  \citet{2015A&A...576A.103B} has shown that the degree of mass removal from tidal interactions is strongly dependent on the location and orbits of galaxies in clusters.  These processes likely do not form very low spin slow rotators, but could reduce the spin of galaxies to match the difference in spin between local spirals and S0s \citep{2021MNRAS.505.2247C}.

Finally, there can be purely gravitational processes internal to galaxies that can gradually dynamically heat stars.    Various simulations \citep[e.g.][]{2016MNRAS.459.3326A,2021MNRAS.503.5826A,2023arXiv230803566Y} find that such heating is possible, for example due to scattering off giant molecular clouds, or the influence of bars and spiral arms.

Given the many above processes, it is vital to be able to narrow down the list to better understand which are more important.  For example, it is particularly interesting to know whether the gas related properties are more or less important than gravitational ones.  \citet{2009MNRAS.393.1324B} considered both colour and optical morphology from Sloan Digital Sky Survey (SDSS) imaging and found that environment vs.\ colour is a stronger trend than environment vs.\ morphology.  This result could point towards gas processes (that change stellar population ages) being more important than dynamical effects \citep[see also][]{2008ApJ...675L..13V,2008MNRAS.387...79V}.

As well as mass and environment, $\lamre$ has also been shown to be a function of stellar population age by  \citet{2018NatAs...2..483V}.  Building on this and other recent results \citep{2017ApJ...844...59B,2021MNRAS.505.3078V,2021ApJ...918...84R}, the aim in our current paper is to combine mass, environment and stellar population parameters to find which are the most important drivers of galaxy spin.  This analysis will make use of the full SAMI Data Release 3 sample \citep{2021MNRAS.505..991C} as well as the detailed environmental data available within the SAMI volume, including the field, groups and clusters from the Galaxy And Mass Assembly (GAMA) Survey \citep{2011MNRAS.416.2640R,2022MNRAS.513..439D} and the SAMI Cluster Survey \citep{2017MNRAS.468.1824O}.

In Section \ref{sec:data} we discuss the data used in our work.  Section \ref{sec:method} describes our statistical methods.  Section \ref{sec:results} presents our main results.  Analysis of simulations is presented in Section \ref{sec:sims} and discussion of the consequences of our results can be found in Section \ref{sec:discussion}.  Our conclusions are presented in Section \ref{sec:conc}.  We assume a cosmology with $\Omega_{\rm m}=0.3$, $\Omega_\Lambda=0.7$ and  $H_0=70$\,\kmsmpc.

\section{SAMI data and measurements}\label{sec:data}

\subsection{SAMI data}

The SAMI instrument \citep{2012MNRAS.421..872C} was deployed at the prime focus of the Anglo-Australian Telescope.  SAMI used 13 imaging fibre bundles \citep[hexabundles;][]{2011OpExpr.19.2649,2014MNRAS.438.869} deployed anywhere across a 1-degree diameter field--of--view.  The hexabundles each contained 61 fibres (each of diameter 1.6 arcsec) that covered a circular area of 15-arcsec diameter on the sky.  The SAMI fibres were fed to the dual--beam AAOmega spectrograph \citep{2006SPIE.6269E..14S}.

The SAMI Galaxy Survey \citep{2012MNRAS.421..872C,2015MNRAS.447.2857B,2021MNRAS.505..991C} observed over 3000 galaxies across a broad range in stellar mass from $\lmstar\sim8$ to $\sim12$, below $z=0.115$.  Target selection was based on SDSS photometry and spectroscopy from the GAMA Survey \citep{2011MNRAS.413..971D}.  As the GAMA fields do not contain massive clusters below $z=0.1$, we separately targeted eight massive clusters to sample galaxy properties in the densest regions \citep{2017MNRAS.468.1824O}.  Stellar masses were estimated using Eq.\ 3 of \citet{2015MNRAS.447.2857B} based on the relation derived for GAMA galaxies by \citet{2011MNRAS.418.1587T}.  

Galaxies were typically observed for 3.5 hours and the data covers the wavelength ranges 3750--5750\,\AA\ and 6300--7400\,\AA\ in the blue and red arms respectively. The spectral resolution of the data is $R=1808$ (at 4800\,\AA) and $R=4304$ (at 6850\,\AA)  \citep{2017ApJ...835..104V,2018MNRAS.481.2299S}.  

The data is reduced using a combination of the \textsc{2dFdr} fibre reduction code \citep{2015ascl.soft05015A} and a purpose built pipeline \citep{2014ascl.soft07006A}.  A detailed description is given by  \cite{2015MNRAS.446.1551S} and \cite{2015MNRAS.446.1567A}, with additions and updates described by \cite{2018MNRAS.481.2299S} and \cite{2021MNRAS.505..991C}.

\subsection{Kinematic measurements}\label{sec:kin}

We use kinematic measurements from SAMI data release 3 \citep{2021MNRAS.505..991C}.  The stellar kinematics are measured using the penalized pixel fitting routine, \ppxf\ \citep{2004PASP..116..138C}, as described in detail by \cite{2017ApJ...835..104V}.  The red and blue arms are fit simultaneously, with the red arm first being convolved to match the spectral resolution in the blue arm.  A Gaussian LOSVD was assumed and optimal templates fit using the MILES stellar library \citep{2006MNRAS.371..703S}. 

\cite{2017ApJ...835..104V} suggest the following quality cuts that we apply to the kinematic maps: signal--to--noise ratio $>3\,$\AA$^{-1}$; $\sigma_{\rm obs} > {\rm FWHM}_{\rm instr}/2 \simeq 35$\,\kms; $V_{\rm error}<30$\,\kms; $\sigma_{\rm error} < 0.1\sigma_{\rm obs} + 25$\,\kms.  The effective radius ($\re$), position angle (PA) and ellipticity of the SAMI galaxies are measured using Multi-Gaussian Expansion \citep[MGE;][]{1994A&A...285..723E,2002MNRAS.333..400C,2009MNRAS.398.1835S}.  Details of the MGE fitting on SAMI galaxies are given by \citet{2021MNRAS.504.5098D}.  $\lamre$ is measured following the procedure described by \cite{2017ApJ...835..104V}, including an aperture correction to 1$\re$ for galaxies where the data does not extend this far \citep{2017MNRAS.472.1272V}. 

Beam-smearing can modify the estimated values of $\lamre$ at the spatial resolution of large multiplexed integral field surveys such as SAMI.  The beam-smearing tends to lower $\lamre$ by converting velocity into velocity dispersion.  To recover the true underlying $\lamre$ we use the corrections described by \citet{2020MNRAS.497.2018H}, with some additional updates given in \citet{2021MNRAS.505.3078V}.  The beam-smearing corrections are based on analysis of simulated galaxies using the {\small SIMSPIN} software \citep{2020PASA...37...16H} and they are a function of $\sigma_{\rm PSF}/\re$, ellipticity and S\'ersic index, where $\sigma_{\rm PSF}$ describes the width of the observational point spread function.  To minimize any residual impact of beam-smearing, we restrict our sample to galaxies where $\sigma_{\rm PSF}/\re<0.5$.  Post-correction, \citet{2020MNRAS.497.2018H} use simulations to show that the difference between true and beam-smear corrected $\lamre$ is very small, with a mean difference of 0.001\,dex and a scatter of 0.026\,dex.  The median measurement uncertainty on $\lamre$ is 0.01\,dex, so the scatter in the beam-smearing correction dominates the uncertainty in $\lamre$.  We note that making the beam-smearing corrections is important for our work, as mean galaxy age varies across the mass-size plane \citep[e.g.][]{2017MNRAS.472.2833S}.  Thus a bias of $\lamre$ with size could lead to a bias of $\lamre$ with age.

\subsection{Stellar population measurements}

As our primary age estimates we use results from the full spectral fitting of SAMI data described by \citet{2022MNRAS.516.2971V}.  To get a representative age for each galaxy we fit to an aperture spectrum where all spaxels within a $1\re$ ellipse (measured using MGE) are summed into a single spectrum \citep{2018MNRAS.481.2299S}.

The full spectral fitting uses \ppxf\ to fit MILES simple stellar population (SSP) models \cite{2015MNRAS.449.1177V} to the $1\re$ aperture spectra.  The blue and red arms are fit simultaneously, with the red arm convolved to the same resolution as the blue (as for the kinematics above).  The models have a metallicity range of $-2.21<[$Fe/H$]<0.4$, an age range of 30 Myr to 14 Gyr and an $\alpha$ abundance of $0.0<[\alpha/$Fe$]<0.4$.  The models make use of the ‘Bag of Stellar Tracks and Isochrones’ models \citep[BaSTI;][]{2004ApJ...612..168P,2006ApJ...642..797P}.

We fit gas emission lines simultaneously with the stellar population models and use a 10th order multiplicative polynomial to correct for small errors in flux calibration.  Uncertainties are estimated from bootstrapping the spectral fits.  Both luminosity-weighted ages ($\agel$) and mass-weighted ages ($\agem$) are derived.  Although not directly shown in this paper, we also test index-based ages derived by \citet{2017MNRAS.472.2833S}.  For further details of age estimates see \citet{2022MNRAS.516.2971V}.

Star formation rate (SFR) estimates are based on SAMI H$\alpha$ emission line maps.  Non-star forming regions are removed based on lying significantly ($>1\sigma$) above the line defined by \citet{2003MNRAS.346.1055K} in the \oiii/H$\beta$ vs. \nii/H$\alpha$ ionization diagnostic diagram \citep{1981PASP...93....5B}.  H$\alpha$ flux is then summed, corrected for extinction [using an average Balmer decrement per galaxy and the \citet{1989ApJ...345..245C} extinction law] and converted to SFR using the relation of \citet{1994ApJ...435...22K}, but corrected to assume a \citet{2003PASP..115..763C} initial mass function.  In detail we may miss some star formation (outside the SAMI aperture, or mixed with AGN emission), but the above approach allows us to provide a SFR for almost all SAMI galaxies.

Although we do not show them in this paper, we have also checked our results for any differences when using SFRs estimated from SED fitting. These are based on the compilation by \citet{2022MNRAS.517.2677R} that uses SFRs based on Galaxy Evolution Explorer (GALEX) SDSS Wide-field Infrared Survey Explorer (WISE) Legacy Catalog version 2 \cite[GSWLC-2;][]{2016ApJS..227....2S,2018ApJ...859...11S}, with the addition of further measurements from GAMA \citep{2020MNRAS.498.5581B,2022MNRAS.513..439D}.  There are no qualitative differences found when using these SED SFRs as opposed to our H$\alpha$ star formation rates.

\subsection{Environmental measurements}

In our analysis we use three different environmental metrics.   The first is the 5th nearest neighbour surface density, $\Sigma_5$.  The method to estimate this value is described in detail by \citet{2017ApJ...844...59B}.  The surface density is defined as $\Sigma_5=5/\pi d^2$, where $d$ is the  projected comoving distance to the 5th nearest galaxy in the density defining population.   The density defining population is taken from the GAMA Survey \citep{2011MNRAS.413..971D} and SAMI Cluster Redshift Survey \citep{2017MNRAS.468.1824O} with a redshift range of $\pm1000$\,\kms\ of the SAMI galaxy redshift.  The limiting magnitude of the density defining sample is an $r$-band absolute magnitude of $-18.6$ mag, but corrected for evolution so that the actual limit is $M_{\rm r}<-18.6-Qz$ with $Q=1.03$ \citep{2017ApJ...844...59B}.  Only galaxies with \texttt{SurfaceDensityFlag}=0 (exact value) or 1 (effective area correction) are included.  Galaxies where insufficient neighbours are found before intersecting with the survey edge are not included.  Median uncertainties on $\lsf$ are 0.063, estimated from the distance to the 4th and 6th nearest neighbour.

As an alternative to local density we use halo properties from the GAMA group catalogue \cite[G3C v10][]{2011MNRAS.416.2640R,2022MNRAS.513..439D}.  We also include equivalent measurements of the SAMI clusters \citep{2017MNRAS.468.1824O}.  The GAMA group catalogue is built using a friend-of-friends methodology that is then calibrated to simulations.  We use two metrics from the group catalogue.  The first is the halo mass $\mhalo$, which is estimated using group velocity dispersion and projected size, calibrated to simulations (See \citealt{2011MNRAS.416.2640R} for details). The second metric is a classification of whether a galaxy is the central of a group, or a satellite.  For this we use the iterative central galaxy, found by iteratively rejecting galaxies furthest from the $r$-band centre of light.  Comparisons to simulations showed this to be the best agreement with the true centre of the halo and it also agrees with the Brightest Cluster/Group  Galaxy (BCG) 95 percent of the time \citep{2011MNRAS.416.2640R}.  From here onwards we call the iterative central galaxy the {\it central} and the other galaxies in the group {\it satellites}.  Although galaxies not in a group are likely the centrals of lower mass haloes (but without satellites being detected in GAMA), we separate these galaxies by labelling them as {\it isolated}.

In the SAMI clusters equivalent central classification was used based on the method presented by \citet{2020ApJ...896...75S}.  This approach selected the most massive galaxy that was less than $0.25R_{200}$ from the cluster centre.  All other galaxies were designated as satellites.  The halo masses for the clusters were based on those published by \citet{2017MNRAS.468.1824O}, but with the a correction factor of $\times1.25$ to bring them to the same mass scale as the GAMA group catalogue [See discussion by \citet{2017MNRAS.468.1824O} for details].  We call central, satellite or isolated labels {\it environmental class} or {\it class} for short.

\begin{table}
	\centering
	\caption{The number of galaxies in our sample, including separately in the cluster and GAMA regions.  We include numbers for both the full sample, and our restricted sample that is limited to $\lmstar>10$ and does not include slow rotators.}
	\label{tab:sample}
	\begin{tabular}{lrrr} 
\hline
 & & GAMA & Cluster\\
Selection & All & regions & regions\\
\hline
Full sample:\\
\hline
$\lmstar>9.5$  &  2090  &  1252  &  838 \\
$\lmstar>9.5$, $\lamre$ &  1646  &  1030  &  616 \\
$\lmstar>9.5$, $\lamre$, Age &  1604  &  996  &  608 \\
$\lmstar>9.5$, $\lamre$, Age, $\Sigma_5$ &  1571  &  975  &  596 \\
$\lmstar>9.5$, $\lamre$, Age, $\mhalo$ &  1270  &  662  &  608 \\
$\lmstar>9.5$, $\lamre$, Age, cent &  399  &  391  &  8 \\
$\lmstar>9.5$, $\lamre$, Age, sat &  887  &  287  &  600 \\
$\lmstar>9.5$, $\lamre$, Age, iso &  318  &  318  &  0 \\
\hline
Restricted sample:\\
\hline
$\lmstar>10$  &  1459  &  893  &  566 \\
$\lmstar>10$, $\lamre$ &  1184  &  751  &  433 \\
$\lmstar>10$, $\lamre$, Age &  1156  &  729  &  427 \\
$\lmstar>10$, $\lamre$, Age, $\Sigma_5$ &  1139  &  718  &  421 \\
$\lmstar>10$, $\lamre$, Age, $\mhalo$ &  902  &  475  &  427 \\
$\lmstar>10$, $\lamre$, Age, cent &  271  &  270  &  1 \\
$\lmstar>10$, $\lamre$, Age, sat &  644  &  218  &  426 \\
$\lmstar>10$, $\lamre$, Age, iso &  241  &  241  &  0 \\
\hline
\end{tabular}
\end{table}

\subsection{Sample size}

Galaxies were selected for analysis based on them having valid values for each parameter of interest.  The total number of galaxies varies slightly, depending on the combination of parameters used.  The number of galaxies used is listed in Table \ref{tab:sample}.  The sample was selected to have $\lmstar>9.5$ on the basis that below this mass stellar kinematic measurements were highly incomplete \citep{2021MNRAS.508.2307V}.  Above this stellar mass limit 1646/2090 (79 percent) galaxies have reliable $\lamre$ measurements.  Modifying our mass limit to $\lmstar>10.0$ changes these numbers to 1413/1696 (83 percent; these numbers are somewhat different from our restricted sample in Table \ref{tab:sample}, as they include slow rotators, see below) and does not qualitatively change our results.  The galaxies not included in our analysis are lost due to the seeing limit and other constraints described in Section \ref{sec:kin}.  Of the galaxies with good $\lamre$, 1604/1646 (97 percent) have good stellar population age measurements.  

Considering environment, the number of galaxies with good $\lamre$, age and $\sigfive$ is 1571/1604 (98 percent).  Galaxies without $\sigfive$ usually lie at the edge of the survey regions so that the distance to the 5th-nearest neighbour cannot be robustly determined.  1270/1604 galaxies have halo mass estimates, with the remainder being almost entirely isolated galaxies that are not in groups.  A small number of groups (16/391) in the GAMA regions do not have group masses due to the measured velocity dispersions being smaller than the velocity uncertainties on GAMA redshifts.  These are typically for low mass groups with low multiplicity (mostly pairs).  Within the GAMA region there are approximately equal numbers of centrals, satellites and isolated galaxies (391, 287 and 318 respectively).  Obviously, in the SAMI clusters, all but one per cluster are satellites.

While most of our analysis is on the full sample (see Table \ref{tab:sample}), we also test the robustness of our results using what we call the {\it restricted sample}.  Our restricted sample is limited to $\lmstar>10.0$ to increase the completeness of kinematic measurements.  In the restricted sample we also remove slow rotators, to allow us to examine whether the slow rotators drive or dominate the relations we see.  We use the slow rotator selection criteria derived for SAMI by \citet{2021MNRAS.505.3078V}, where slow rotators are classed as objects with
\begin{equation}
  \lamre =\lambda_{\rm R_e,start} + \epsilon_e/4,\,\,\,\,\,\, {\rm and}\,\,\,\,\,\,  \epsilon_e < 0.35+\frac{\lambda_{\rm R_e,start}}{1.538}.
\end{equation}
$\epsilon_e$ is the ellipticity (measured using multi-Gaussian expansion) within 1 $R_e$ and for SAMI $\lambda_{\rm R_e,start}=0.16$.  Using  lower $\lambda_{\rm R_e,start}=0.08$ proposed by \citet{2016ARA&A..54..597C} does not have a qualitative impact on our conclusions.  The number of objects in the restricted sample is shown in Table \ref{tab:sample}.

\begin{table}
	\centering
	\caption{The results of our full and partial correlation analysis between 4 parameters: $\lmstarnu$, $\lsfnu$, $\lamre$ and an age proxy.  The age proxy is either light-weighted age ($\agel$), mass-weighted age ($\agem$) or $\lssfrnu$.  For each pair of parameters $(A,B)$ we list the correlation coefficient ($r$) and probability of the null hypothesis ($P$-value) for the full correlation analysis (just correlating the two parameters) and the partial correlation analysis (accounting for the correlations between other variables).  Correlations that are {\it not} significant ($P$-value$>0.01$) are highlighted in bold.}
	\label{tab:corr_four_full}
	\begin{tabular}{ccrrrr} 
\hline
& & \multicolumn{2}{c}{full correlation} &\multicolumn{2}{c}{partial correlation}\\
A & B & $r$ & $P$-value & $r$ & $P$-value \\
\hline
 \multicolumn{3}{l}{Light-weighted age:}\\
\hline
$\lmstarnu$ & $\agel$ &  0.515 & 2.37e-107 &   0.453 & 2.41e-80 \\
$\lmstarnu$ & $\lsfnu$ &  0.100 & 7.38e-05 &  -0.117 & 3.68e-06 \\
$\lmstarnu$ & $\lamre$ & -0.294 & 9.62e-33 &   0.006 & {\bf 8.12e-01} \\
$\agel$ & $\lsfnu$ &  0.373 & 4.02e-53 &   0.339 & 1.77e-43 \\
$\agel$ & $\lamre$ & -0.576 & 3.00e-139 &  -0.494 & 1.99e-97 \\
$\lsfnu$ & $\lamre$ & -0.199 & 1.99e-15 &   0.022 & {\bf 3.83e-01} \\
\hline
\multicolumn{3}{l}{mass-weighted age:}\\
\hline
$\lmstarnu$ & $\agem$ &  0.576 & 8.12e-140 &   0.526 & 1.61e-112 \\
$\lmstarnu$ & $\lsfnu$ &  0.100 & 7.38e-05 &  -0.087 & 5.40e-04 \\
$\lmstarnu$ & $\lamre$ & -0.294 & 9.62e-33 &  -0.081 & 1.41e-03 \\
$\agem$ & $\lsfnu$ &  0.282 & 4.47e-30 &   0.235 & 3.65e-21 \\
$\agem$ & $\lamre$ & -0.417 & 4.13e-67 &  -0.283 & 2.83e-30 \\
$\lsfnu$ & $\lamre$ & -0.199 & 1.99e-15 &  -0.099 & 8.19e-05 \\
\hline
\multicolumn{3}{l}{Specific star formation rate:}\\
\hline
$\lmstarnu$ & $\lssfrnu$ & -0.367 & 5.10e-52 &  -0.250 & 3.13e-24 \\
$\lmstarnu$ & $\lsfnu$ &  0.105 & 2.71e-05 &  -0.059 & {\bf 1.80e-02} \\
$\lmstarnu$ & $\lamre$ & -0.304 & 1.86e-35 &  -0.122 & 1.08e-06 \\
$\lssfrnu$ & $\lsfnu$ & -0.437 & 2.13e-75 &  -0.401 & 1.61e-62 \\
$\lssfrnu$ & $\lamre$ &  0.567 & 3.22e-136 &   0.491 & 1.65e-97 \\
$\lsfnu$ & $\lamre$ & -0.203 & 2.99e-16 &   0.053 & {\bf 3.47e-02} \\
\hline
\end{tabular}
\end{table}

\begin{figure*}
	\includegraphics[trim=0mm 5mm 30mm 30mm, clip, width=18.0cm]{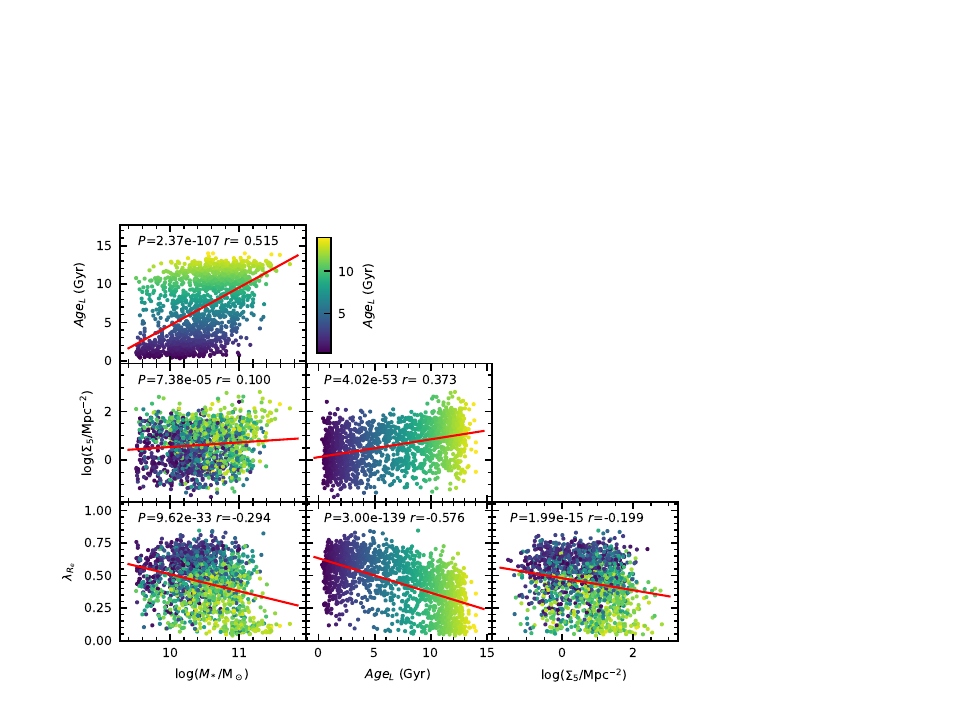}
    \caption{The correlation between each of our 4 main variables, $\lmstarnu$, $\lsfnu$, $\lamre$ and age (for this example we choose $\agel$).  Here the correlations are shown without accounting for correlations in other parameters.  The points are colour coded by $\agel$.  The red line shows the linear correlation between the two variables and in each panel the $P$-value of the null hypothesis that they are uncorrelated is given.}  
    \label{fig:lam_agel_mass_sigma5_corner}
\end{figure*}

\subsection{EAGLE simulation data}

We will compare our results from the SAMI Galaxy Survey to the EAGLE simulations \citep{2015MNRAS.446..521S,2015MNRAS.450.1937C,2017MNRAS.464.3850L}. Our primary aim here is to see if the simulations show similar trends to the data and to predict trends to higher redshift.  For more detailed comparisons between SAMI data and simulations see \citet{2019MNRAS.484..869V}.  We make use of the EAGLE reference model (Ref-L100N1504), using derived quantities calculated as described by \citet{2018MNRAS.476.4327L}.  We adopt a lower limit on the stellar mass of simulated galaxies of $\lmstar=10$ to make sure that we are not impacted by resolution effects, including when we mass match to the SAMI data, so we only match in stellar mass above this limit.  $\lamre$ is calculated by deriving $r$-band weighted kinematics from stellar particles projected onto a 2D grid of size 1.5\,kpc (proper distance).  The projection is done both at a random inclination and for an edge-on view, although we will use the random inclination to directly compare to our data.  We also estimate $\lamre$ weighted by stellar mass, rather than $r$-band light.  This allows us to examine the impact that light vs.\ mass weighting has on the simulated properties.  Mean stellar population age is calculated both mass-weighted and light-weighted ($r$-band), with the $r$-band luminosity estimated using \citet{2003MNRAS.344.1000B} stellar population synthesis models and a \citet{2003PASP..115..763C} initial mass function.  From EAGLE we also calculate a projected 5th nearest neighbour density as an estimate of environment using a similar approach to that used for the SAMI sample.  The density defining population is selected to have $M_{\rm r}>-18.6$, the same as the SAMI $M_{\rm r}$ limit.  The velocity range used to estimate the density was $\pm1000$\,\kms. Halo masses and whether a galaxy is a central or satellite in its halo were also extracted from the simulations.

\section{Methods}\label{sec:method}

In this paper we will primarily use two statistical approaches.  The first is a linear correlation analysis, including partial correlations.  For this analysis we use the \textsc{Pingouin} \citep{Vallat2018} and \textsc{Statsmodels} \citep{seabold2010statsmodels} packages.  As a first step we calculate the Pearson correlation coefficient for each pair of variables.  To find the dominant drivers of correlations we carry out a partial correlation analysis \citep[e.g.][]{1982MNRAS.199.1119M,2002MNRAS.337..275C,2022MNRAS.512.1765O}.  This analysis calculates the correlation coefficients between two variables, while taking into account the correlations with other variables.  This approach effectively finds the residual correlation between parameters $A$ and $B$ after removing the correlation with the other two parameters (that we label $X_2$), i.e.\ $\Delta A|X_2$ vs.\ $\Delta B|X_2$.  A specific example would be the correlation between $\Delta\lamre|\lmstarnu,\lsfnu$ and $\Delta\agel|\lmstarnu,\lsfnu$.  In this case $X_2 = \lmstarnu,\lsfnu$.  We also consider the case of completely ignoring variables in the analysis, such as looking at the correlation between $\lamre$ and $\lsfnu$ while only controlling for age, and ignoring mass. This approach allows us to gain further insight into the main drivers of the correlations found.  In general we will treat correlations with a $P$-value of $<0.01$ as significant.

Our second approach will be to fit the $\lamre$--age relation and then examine whether there are residual differences as a function of environment away from the mean relation.  We will see below that the mean trend of $\lamre$ vs.\ age is well approximated by a sigmoid function of the form
\begin{equation}
\lamre =  \lambda_0 - \frac{\lambda_0-\lambda_1}{1 + e^{-k(x-x_0)}}.
\label{eq:sigmoid}
\end{equation}
The values $\lambda_0$ and $\lambda_1$ are the asymptotic values of $\lamre$ at young and old age respectively.  The variable $x$ is an age measurement, while $x_0$ and $k$ determine the location and sharpness of the transition.  To quantify the significance of any difference as a function of environment we will remove the $\lamre$ vs.\ age trend by subtracting the best fitting sigmoid function from the individual $\lamre$ values for each galaxy based on their measured ages, to derive a $\Delta\lamre$.  We then use a Kolmogorov–Smirnov (K-S) test to compare the distributions of $\Delta\lamre$ between samples with different environmental measurements.

\section{The relationship between spin age and environment in SAMI galaxies}\label{sec:results}

\subsection{Correlation analysis}\label{sec:correlation}

\begin{figure*}
	\includegraphics[trim=0mm 5mm 30mm 30mm, clip, width=18.0cm]{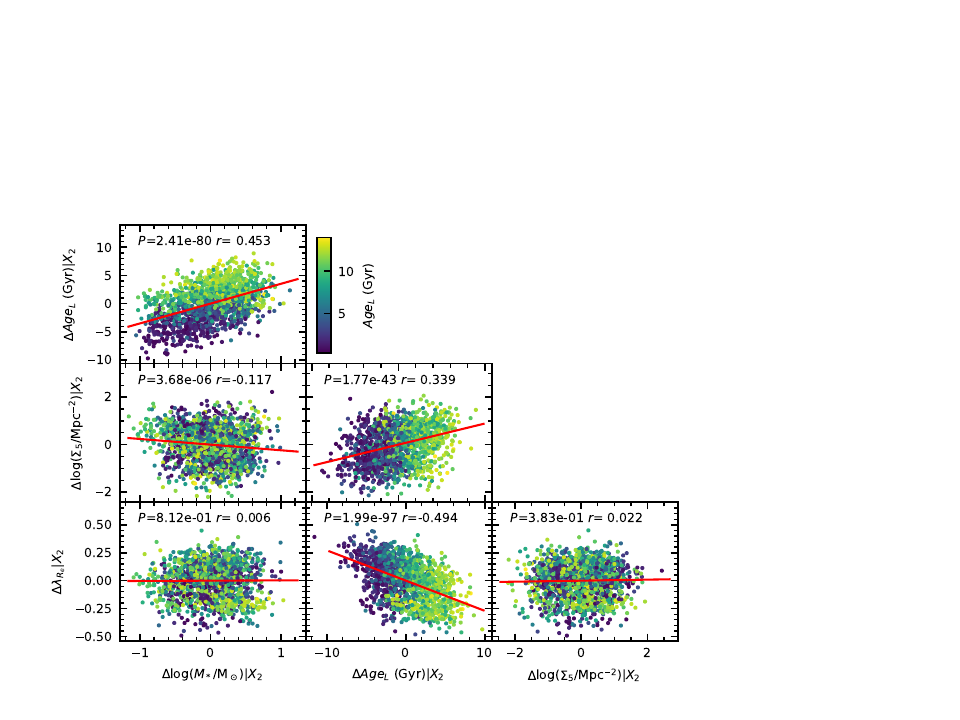}
    \caption{The correlation between each of our 4 main variables, $\lmstarnu$, $\lsfnu$, $\lamre$ and age (for this example we choose $\agel$).  The data is identical to Fig.\ \ref{fig:lam_agel_mass_sigma5_corner}, but in this case we are showing the residual correlation between two variables once the correlation with the other two variables (labelled $X_2$) has been accounted for.  The points are colour-coded by $\agel$.  The red line shows the linear correlation between the two variables and in each panel the $P$-value of the null hypothesis that they are uncorrelated is given.}  
    \label{fig:lam_agel_mass_sigma5_corner_res}
\end{figure*}

To begin our investigation we look at the connection between $\lamre$, $\lmstarnu$, $\lsfnu$ and age by carrying out a correlation analysis using the approach discussed in Section \ref{sec:method}.  This will generally use the full sample with $\lmstar>9.5$ that contains 1571 galaxies when using $\lsfnu$ as our environmental metric.  Correlations for each of our three different age proxies are discussed below.  

\subsubsection{Correlations with light-weighted age}

First we consider light-weighted age from full spectral fitting, $\agel$, as our age measurement.  Although not shown, we note that results are qualitatively similar when using Lick-index-based ages \citep{2017MNRAS.472.2833S}.  To begin with we show the correlations within our full sample between each variable {\it without} taking other correlations into account (Fig.\ \ref{fig:lam_agel_mass_sigma5_corner} and Table \ref{tab:corr_four_full}).  We find various well known trends, such as the strong correlation between mass and age, $\lamre$ and age etc.  In fact, there are significant correlations between all variables, although the one between $\lsfnu$ and $\lmstarnu$ is the weakest.  It should be noted that in our correlation analysis we are assuming a linear relationship between our variables (red lines in  Fig.\ \ref{fig:lam_agel_mass_sigma5_corner}), which is unlikely to be true in detail.  However, a Spearman rank correlation shows the same qualitative results.

In Fig.\ \ref{fig:lam_agel_mass_sigma5_corner_res} we present the correlations between our variables again, but this time accounting for correlations in other parameters (see also Table \ref{tab:corr_four_full}).  When removing the correlation due to other variables we still find highly significant correlations between $\agel$ and all three of the other variables.  For example, we find the correlation between $\agel$ and $\lamre$ is highly significant ($r=-0.494$, $P=1.99e-97$), even after we control for $\lmstarnu$ and $\lsfnu$.  In contrast, if $\agel$ is one of the controlled variables, the remaining correlations with $\lamre$ are not significant.  The partial correlation of $\lamre$ and $\lsfnu$ is not significant ($r=0.022$, $P=3.83e-01$) and neither is the partial correlation of $\lamre$ and $\lmstarnu$ ($r=0.006$, $P=8.12e-01$).  These results seem to suggest that light-weighted age is a more fundamental parameter for defining the spin of a galaxy than either mass or environment.

The partial correlation between $\lmstarnu$ and $\lsfnu$ is significant (although weaker than other significant correlations) and opposite in sign to the full correlation of these variables.    Given the sense of the partial correlation infers higher mass galaxies in lower density environments, this is somewhat counter intuitive.  However, this correlation is no longer significant when we separately examine the $\lmstarnu$ vs.\ $\lsfnu$ correlation in the GAMA and cluster regions.  The significant partial correlation in the full sample  is therefore driven by GAMA and cluster galaxies lying in somewhat different parts of the $\lmstarnu$ vs.\ $\lsfnu$ plane [e.g. see Fig. 17 of \citet{2021MNRAS.505..991C}].  This does not influence our overall conclusions (see Sections \ref{sec:full_sample} and \ref{sec:mass}).    

To further investigate the above findings we repeat the partial correlation analysis, but remove one variable from the analysis, so that only three variables are considered at a time.  This approach allows us to get a clearer picture of which variables are most important.  When we ignore $\agel$ there are highly significant remaining partial correlations between $\lmstarnu$ and $\lamre$, and $\lsfnu$ and $\lamre$.  A single parameter (either mass or environment) is not sufficient to understand the trends in $\lamre$. In contrast, when we ignore $\lmstarnu$ the partial correlation between $\lamre$ and $\lsfnu$ is not significant, controlling for only $\agel$.  What is more, ignoring $\lsfnu$, the partial correlation between $\lamre$ and $\lmstarnu$ is not significant once $\agel$ is controlled for.  The implication is that the relationship between $\agel$ and $\lamre$ is sufficient to drive the $\lamre - \lsfnu$ and  $\lamre - \lmstarnu$ relationships.

To confirm that the above results are robust, we repeat the analysis, but with different sub-samples.  The detailed correlation analysis results for these tests are listed in Appendix \ref{app:a}.  When limiting our analysis to the restricted sample ($\lmstar>10$, no slow rotators) we find the same key conclusion; once we control for $\agel$, there is no significant correlation between $\lamre$ and $\lsfnu$ or $\lmstarnu$.  However, the correlation between $\agel$ and $\lamre$ is still strong ($r\simeq-0.5$).  Likewise, when we limit the analysis to the restricted sample, but separate the GAMA and cluster regions we also find this same overall conclusion.  

In summary, light-weighted age ($\agel$) is the variable that best correlates with $\lamre$.  Once the correlation with $\agel$ is controlled for, there is no significant residual correlation with $\lsfnu$ or $\lmstarnu$ in our full sample ($\lmstar>9.5$, including slow rotators).  This result is not driven by slow rotators, as a similar result is found in a more restricted sample that does not include slow rotators.

\subsubsection{Correlations with mass-weighted age}\label{sec:corrmasswt}

When we use mass-weighted age, $\agem$, we find somewhat different results (Table \ref{tab:corr_four_full}).  As for $\agel$,  the full correlations using $\agem$ are significant in all cases.  However, while the partial correlations between $\agem$ and other variables are strongest, all the other partial correlations are also significant.  $\agem$ is the variable with the strongest partial correlation with $\lamre$ ($r=-0.283$, $P=2.83e-30$), but the partial correlations of both $\lmstarnu$ and 
$\lsfnu$ with $\lamre$ are also significant ($r=-0.081$. $P=1.41e-03$ and $r=-0.099$ $P=8.19e-05$ respectively).  The partial correlations when ignoring one variable again highlight that the strongest correlations are with $\agem$, but unlike for $\agel$, all correlations remain significant.  The strength of the $\agem-\lamre$ partial correlation ($r=-0.283$) is less than the $\agel-\lamre$ partial correlation ($r=-0.494$).  This may be because $\agel$ is more more directly related to $\lamre$, or could be due to mass weighted ages being more difficult to measure, with larger uncertainties.  We explore this using the EAGLE simulations in Section \ref{sec:lowzsims} below.

Testing the restricted sample ($\lmstar>10$, no slow rotators), all partial  correlations remain significant, with the exception of that between $\lmstarnu$ and $\lsfnu$ (See Appendix \ref{app:a}).  However, the strongest correlations remain between $\agem$ and $\lmstarnu$ or $\lamre$.    The weakest significant correlations are between $\lamre$ and $\lmstarnu$, and $\lamre$ and $\lsfnu$.  When analysing the restricted sample within the GAMA and cluster subsets independently, the same broad picture emerges.  However, in the GAMA regions the weak partial correlation between $\lamre$ and $\lsfnu$ is no longer significant.  For the cluster regions we also find the $\lamre-\lsfnu$ partial correlation is no longer significant, but neither is the  $\lmstarnu-\lamre$ case.  Interestingly, the $\agem-\lsfnu$ correlation is also not significant in the clusters, likely due to a reduced dynamic range of environment and age for cluster galaxies.

In summary, when using $\agem$ as our age estimate, age is still the variable that correlates most strongly with $\lamre$.  However, the correlation is not as strong as for $\agel$, and correlations with other variables are required to explain the distribution of $\lamre$.

\begin{figure*}
	\includegraphics[trim=8mm 4mm 45mm 15mm, clip, height=5.2cm]{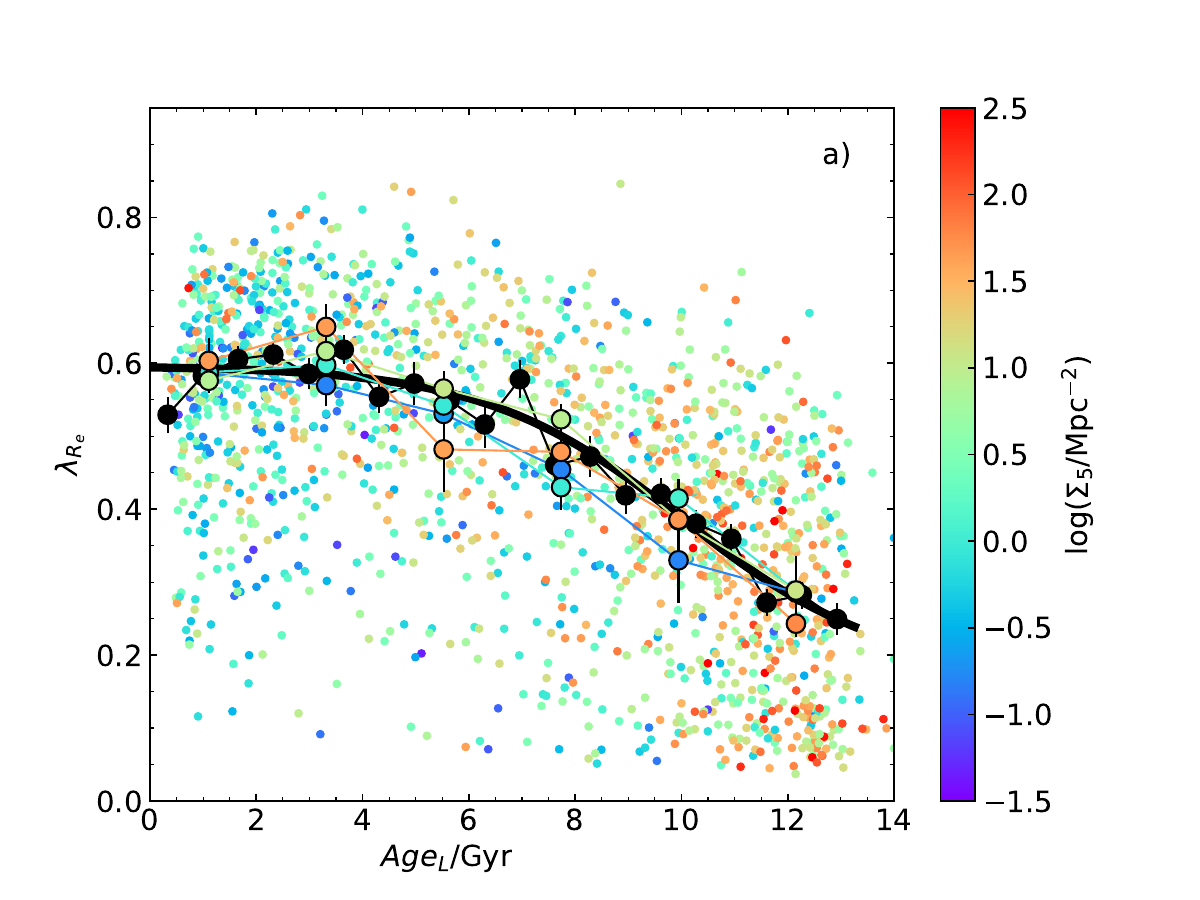}\includegraphics[trim=15mm 4mm 45mm 15mm,clip,height=5.2cm]{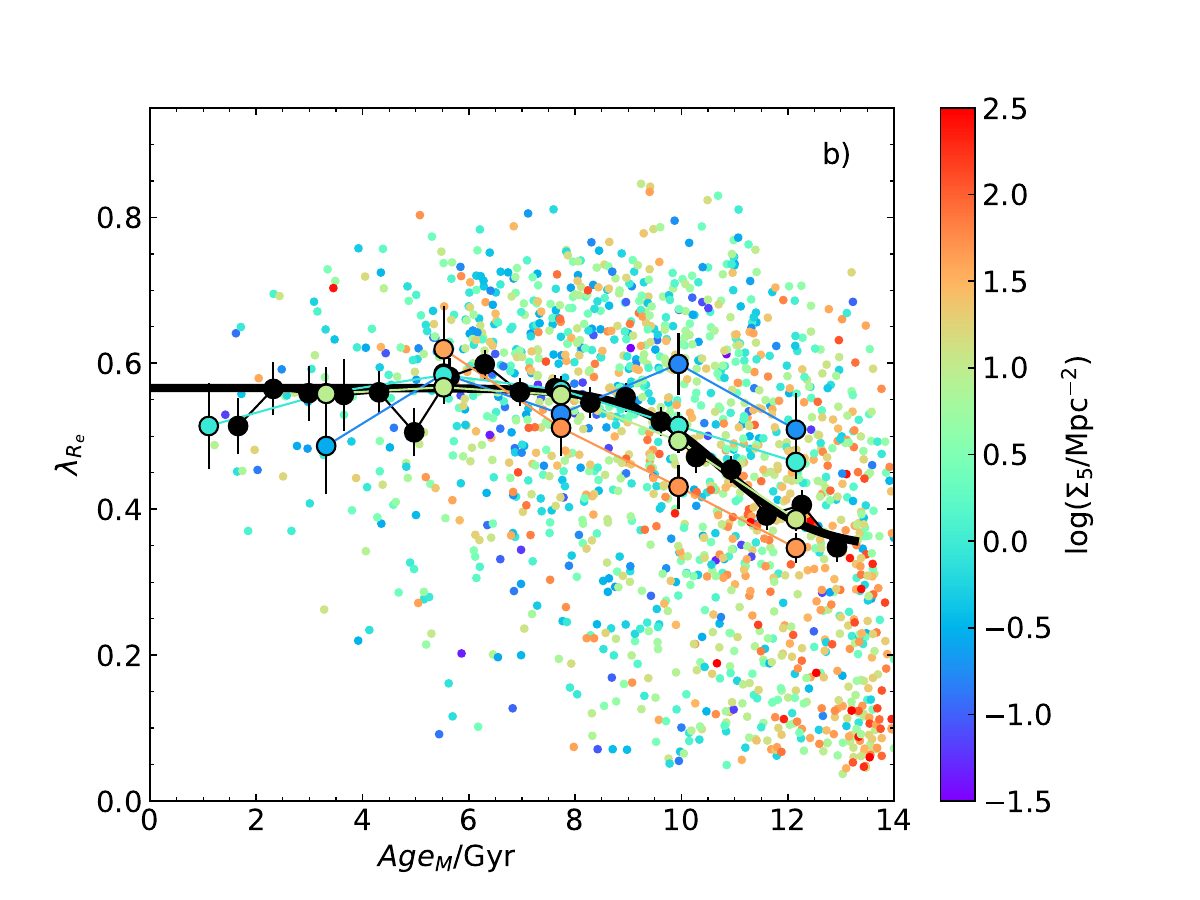}\includegraphics[trim=15mm 4mm 5mm 15mm,clip,height=5.2cm]{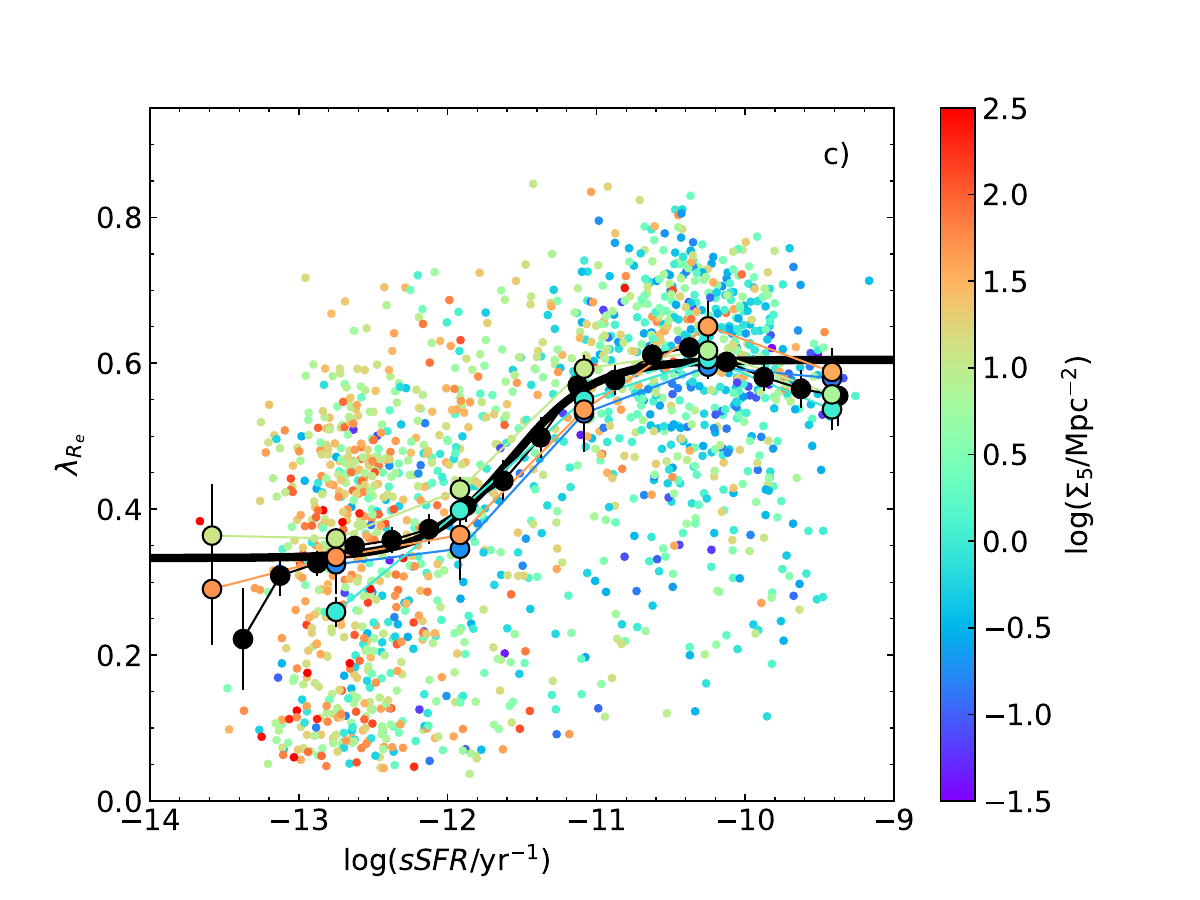}
	\includegraphics[trim=8mm 4mm 45mm 15mm,clip,height=5.2cm]{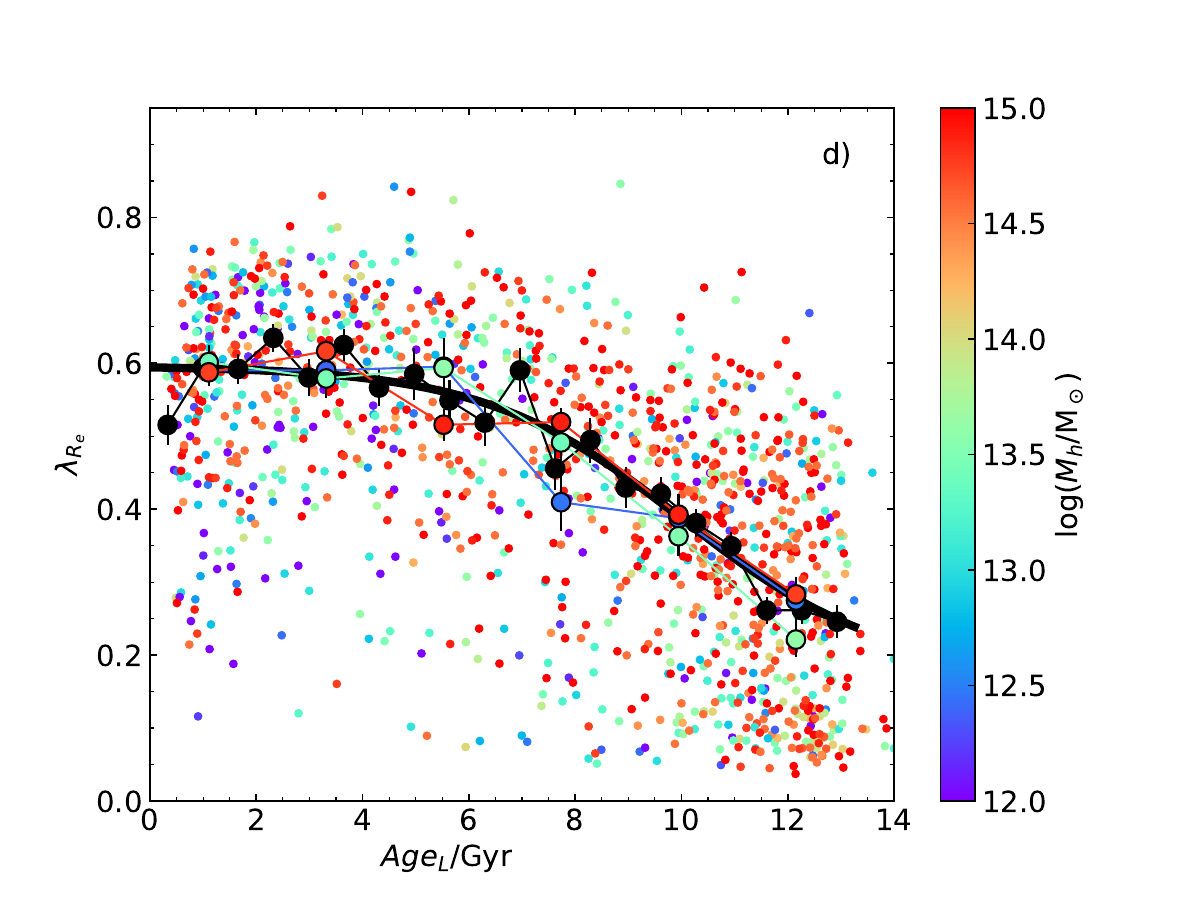}\includegraphics[trim=15mm 4mm 45mm 15mm,clip,height=5.2cm]{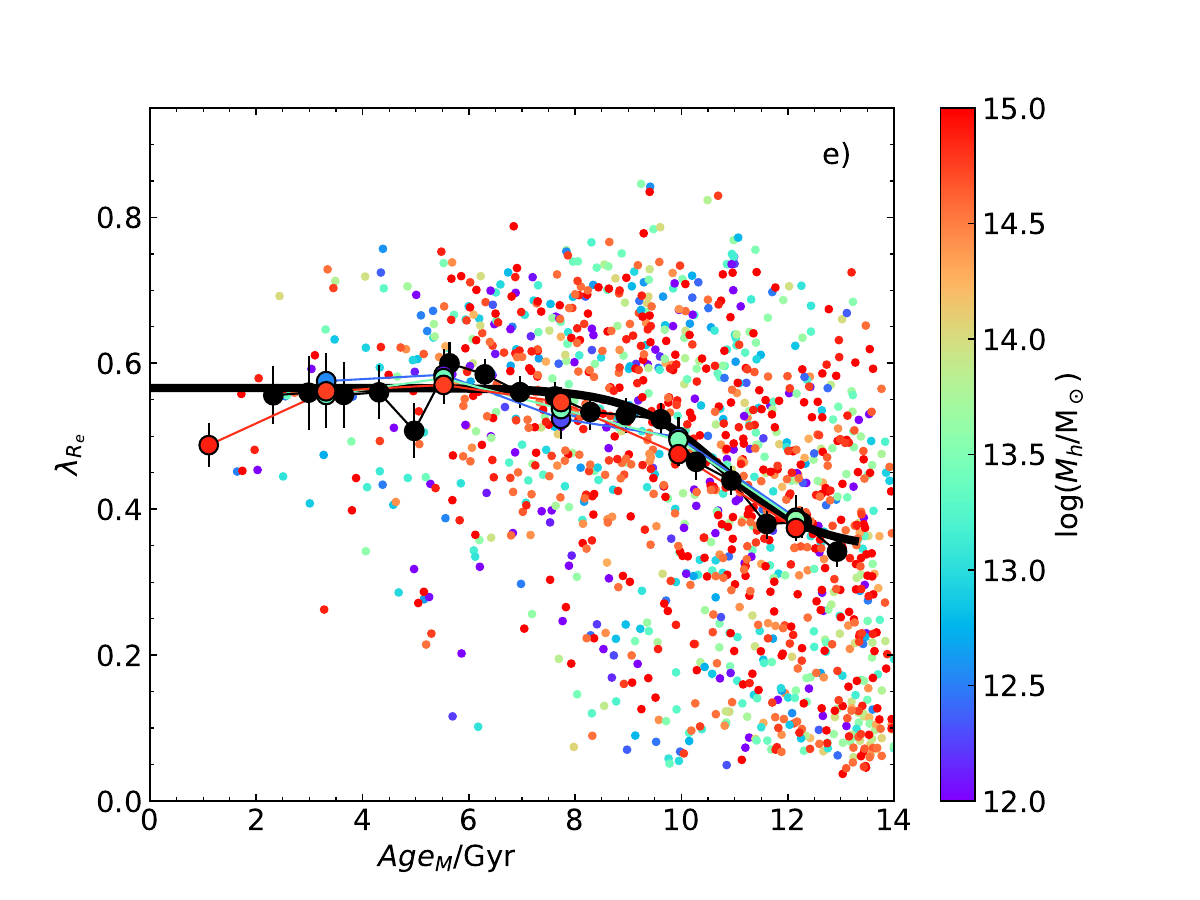}\includegraphics[trim=15mm 4mm 5mm 15mm,clip,height=5.2cm]{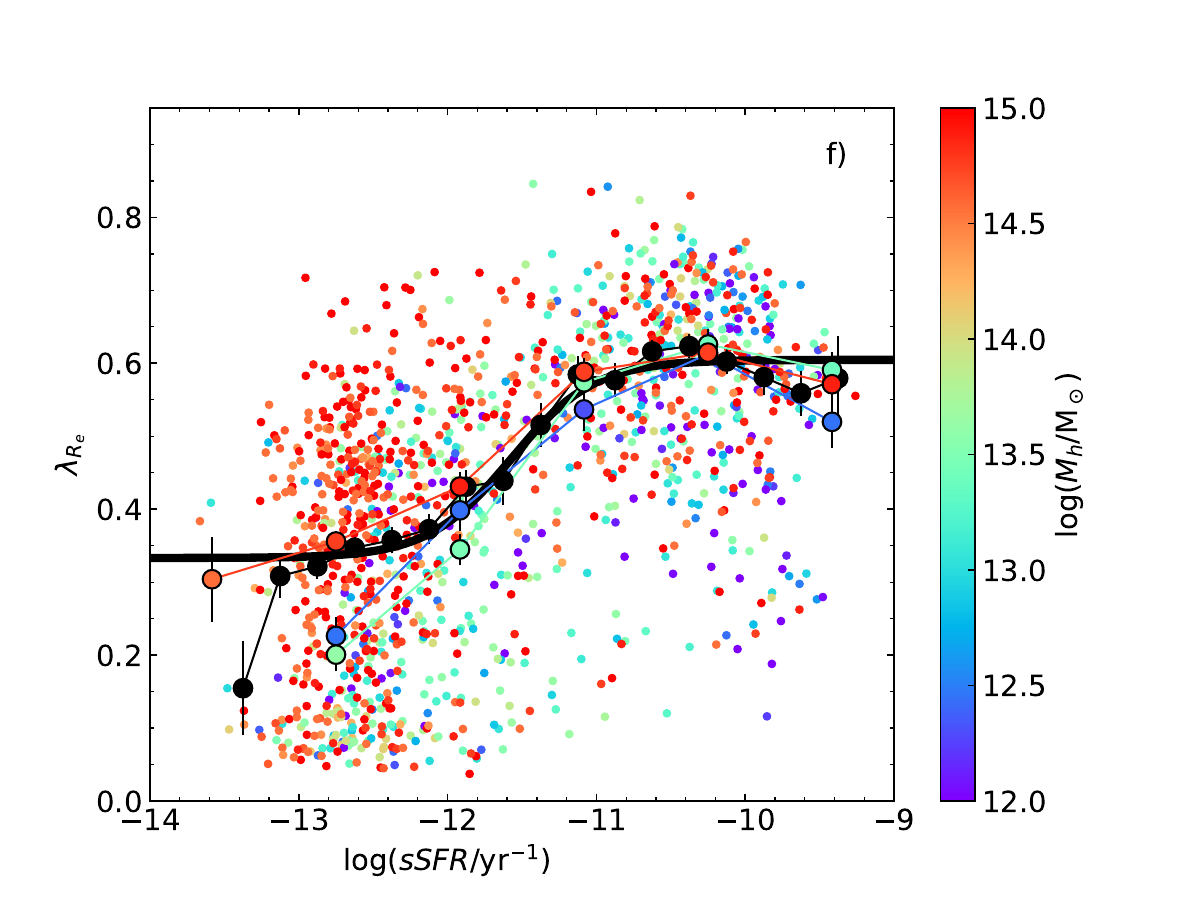}
	\includegraphics[trim=8mm 4mm 45mm 15mm, clip, height=5.2cm]{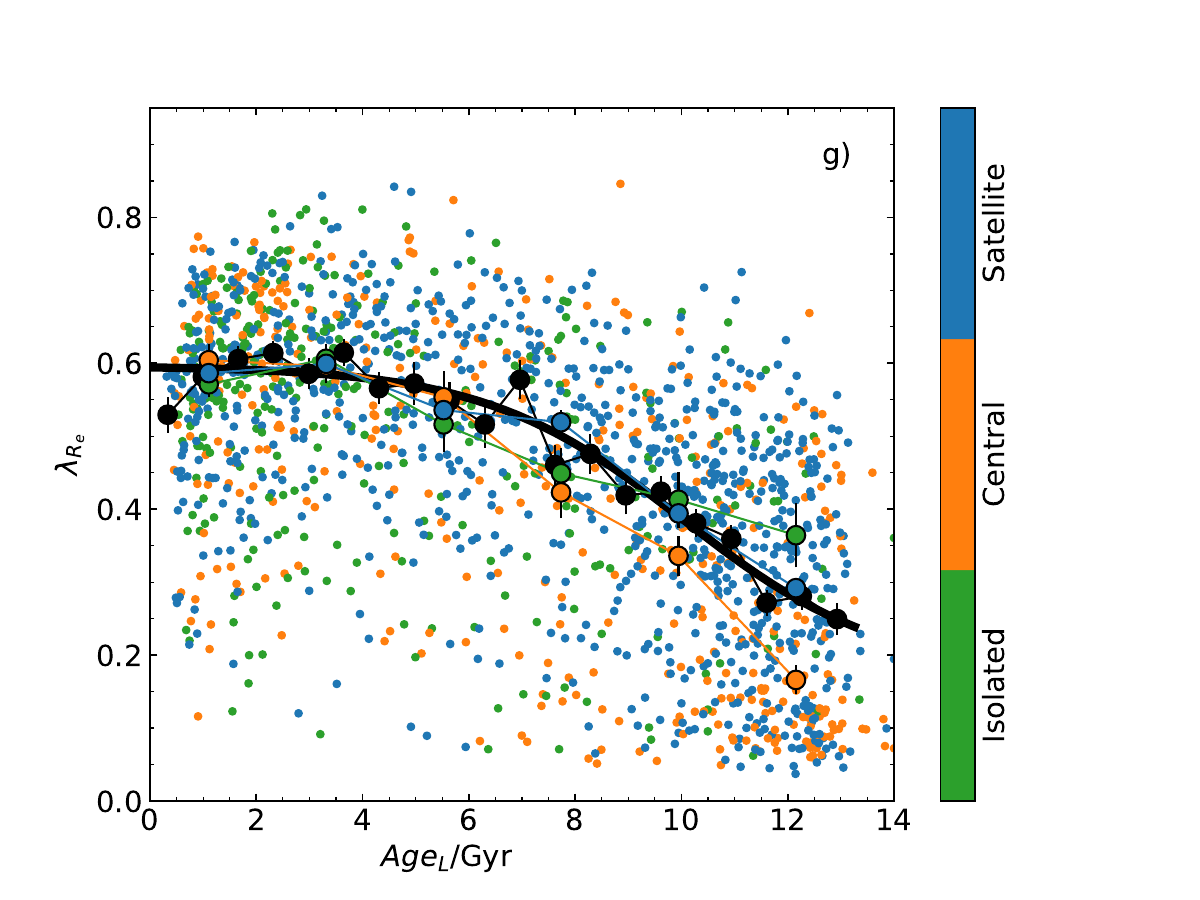}\includegraphics[trim=15mm 4mm 45mm 15mm,clip,height=5.2cm]{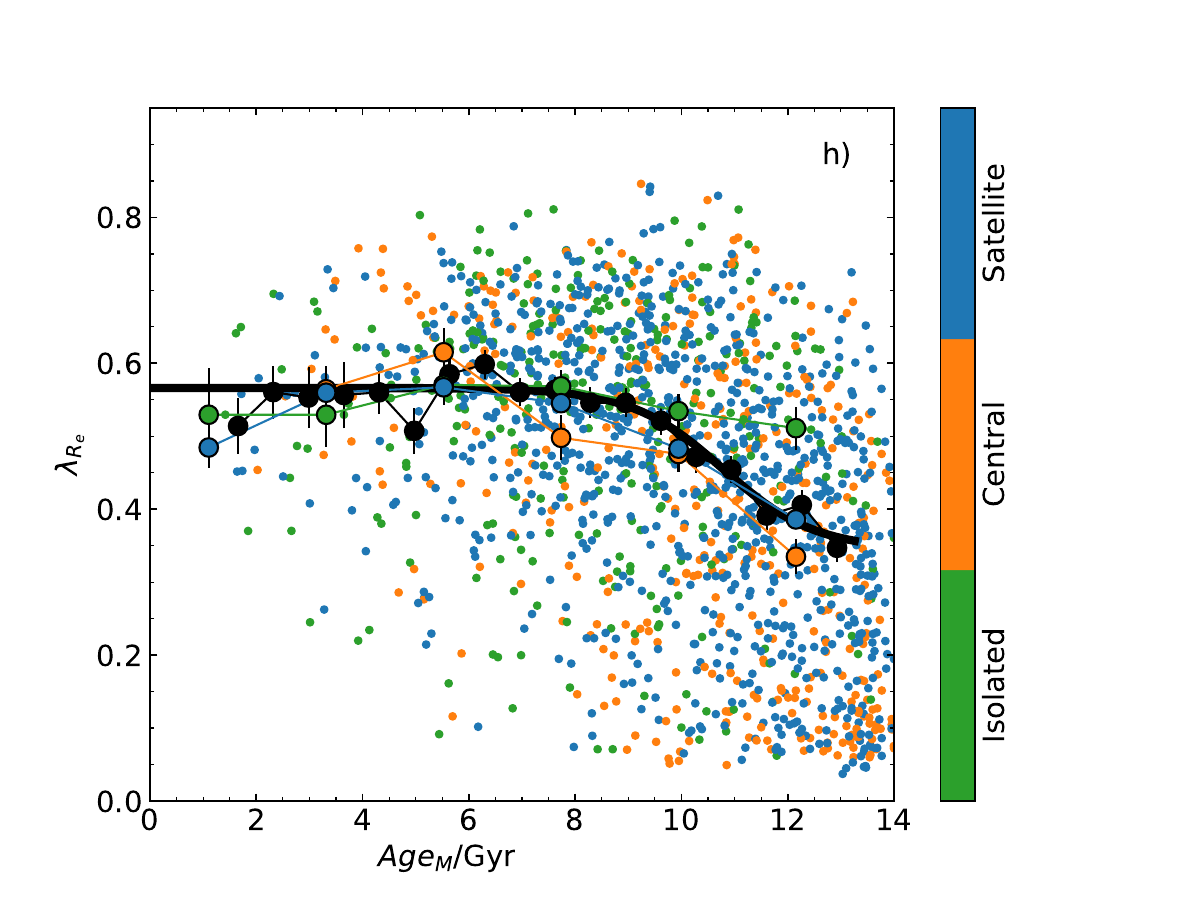}\includegraphics[trim=15mm 4mm 5mm 15mm,clip,height=5.2cm]{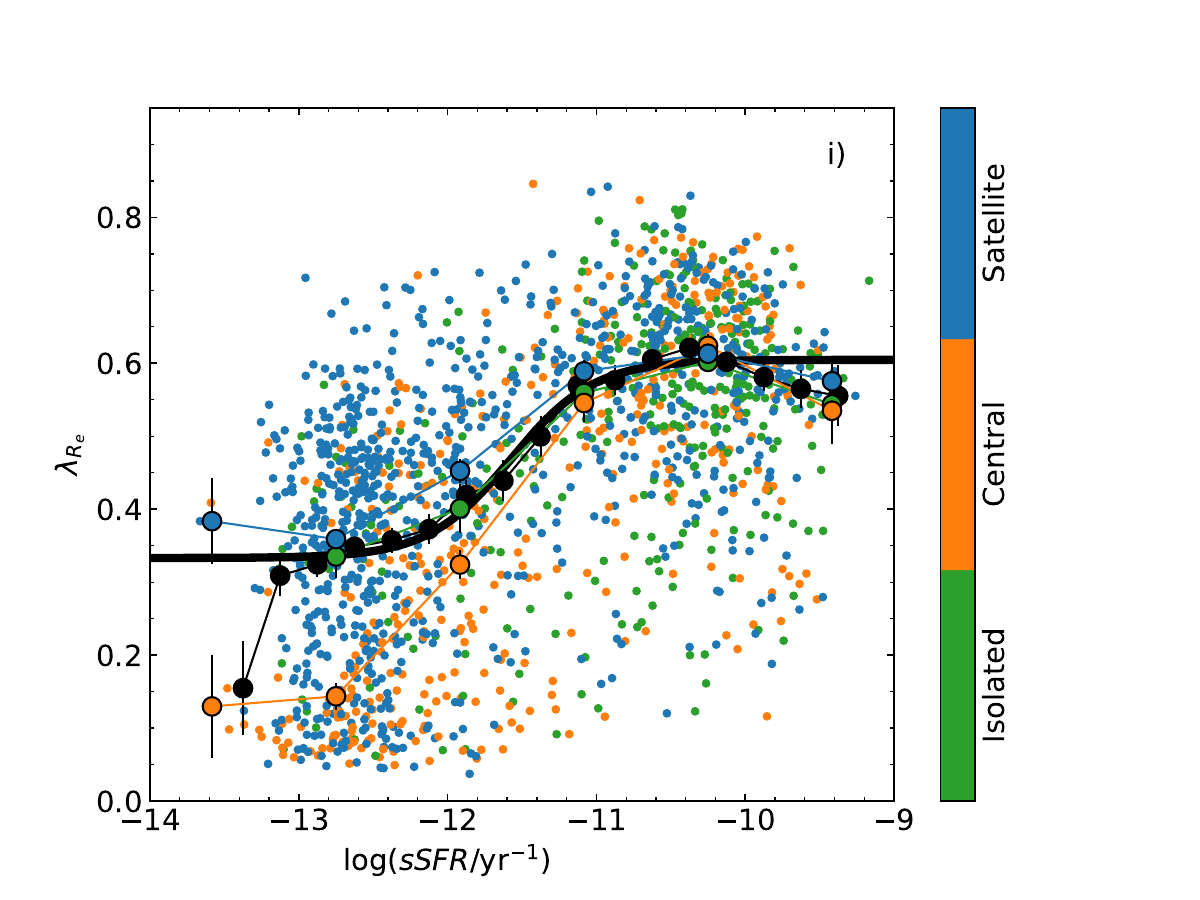}
    \caption{The relation between $\lamre$ and age proxy for SAMI galaxies, colour-coded by environment using $\Sigma_5$ (a, b, c), halo mass (d, e, f), and halo location (g, h, i).  The left-hand side (a, d, g) uses light-weighted age, $\agel$, the central panels (b, e, h) use mass-weighted age, $\agem$ and the right-hand side (c, f, i) uses sSFR.  In each panel the large black points show the median spin in age bins.  These are fit with a sigmoid function (thick black line).  The large coloured points are the median spin in age bins, but separated into different intervals in environment, using the same colour coding as the smaller points.  Error bars show the uncertainty on the median, calculated using the 68th percentile width of the distribution.}
    \label{fig:lam_age_sigma5}
\end{figure*}

\begin{table*}
\centering
	\caption{Results of the sigmoid fit to the median age vs. $\lamre$ relations for different age proxies (see Eq.\ \ref{eq:sigmoid}).  We list the results from SAMI and from EAGLE (stellar mass matched to SAMI).  We also include the RMS scatter of galaxies around the median fitted relation.}
	\label{tab:sigmoid_fit}
	\begin{tabular}{lccccc} 

\hline
Age type & $\lambda_0$ & $\lambda_1$ & $x_0$ & $k$ & RMS\\
\hline
SAMI:\\
\hline
$\agel$ & $0.5965\pm0.0002$ & $0.1742\pm0.0083$ & $10.04\pm0.84$ & $0.53\pm0.03$ & 0.153\\
$\agem$ & $0.5663\pm0.0001$ & $0.3447\pm0.0011$ & $10.78\pm0.14$ & $1.15\pm0.14$ & 0.170\\
$\log(sSFR)$ & $0.6048\pm0.0001$ & $0.3330\pm0.0001$ & $-11.57\pm0.01$ & $-3.58\pm0.72$ & 0.154\\
\hline
EAGLE:\\
\hline
$\agel$ & $0.5240\pm0.0006$ & $0.1700\pm0.0037$ & $ 9.18\pm0.18$ & $1.03\pm0.15$ & 0.196\\
$\agem$ & $0.4723\pm0.0002$ & $0.2019\pm0.0022$ & $10.10\pm0.08$ & $1.64\pm0.34$ & 0.205\\
$\log(sSFR)$ & $0.6133\pm0.0011$ & $0.2246\pm0.0002$ & $-10.77\pm0.01$ & $-4.03\pm1.12$ & 0.169\\
\hline
    \end{tabular}
\end{table*}

\subsubsection{Correlations with specific star formation rate}

Our third age proxy is specific star formation rate (sSFR).  The results of our correlation analysis on the full sample using $\lssfrnu$ are listed in Table \ref{tab:corr_four_full}.  As with all the other age proxies, the full correlations are significant between all variables (3rd and 4th columns in Table \ref{tab:corr_four_full}).  The strongest full correlation found is between $\lssfrnu$ and $\lamre$ ($r=0.491$, $P=1.95e-97$), followed by the $\lssfrnu-\lsfnu$ ($r=-0.401$, $P=1.61e-62$) correlation then the $\lssfrnu-\lmstarnu$ correlation ($r=-0.250$, $P=3.13e-24$). 

For the partial correlation analysis including $\lssfrnu$, the $\lssfrnu-\lamre$ and $\lssfrnu-\lsfnu$ correlations remain the strongest.  The $\lamre-\lmstarnu$ correlation is reduced, but still significant, while the $\lamre-\lsfnu$ correlation ceases to be significant.  These results are qualitatively similar to the trends we see above when using $\agel$.

The partial correlation analysis, ignoring one variable shows similar trends to those above.  $\lssfrnu$ consistently shows the strongest correlations with other variables.  If we ignore $\lmstarnu$, then control only for $\lssfrnu$, there is no residual correlation between $\lamre$ and $\lsfnu$.

Repeating our analysis with different sub-samples we find the same overall trends (details in Appendix \ref{app:a}).  Using the restricted sample, $\lssfrnu$ consistently shows the strongest partial correlations and there is no correlation between $\lamre$ and $\lsfnu$ once the other variables are controlled for.  If we separate the GAMA and cluster regions, the strongest correlation (full and partial) in each is between $\lssfrnu$ and $\lamre$.

Although we don't present the results in this paper, we investigate if SED derived SFRs \citep{2022MNRAS.517.2677R} agree with our results using H$\alpha$.  The SED SFRs  provide the same qualitative picture, with sSFR being much more strongly correlated with $\lamre$ than $\lmstarnu$ or $\lsfnu$.  The SED SFRs show a slightly reduced partial correlation ($r=0.432$) compare to H$\alpha$ ($r=0.491$).    The SED SFRs also show a weak partial correlation between $\lamre$ and $\lmstarnu$, and no partial correlation between $\lamre$ and $\lsfnu$, consistent with the H$\alpha$ measurements.

\subsubsection{Summary of correlation analysis}

There is a consistent pattern across all of the correlation analyses discussed above.  In every case, age (or sSFR) correlates most strongly with the other variables and these correlations are always significant.  The correlations with age or sSFR remain significant if we control for other variables via a partial correlation analysis.  In contrast,  $\lamre$ does not always show significant correlations.  Once the strong correlation between $\lamre$ and age or sSFR is accounted for, the remaining residual correlations between $\lamre$ and $\lsfnu$ or $\lmstarnu$ are weak or in many cases insignificant.

The two age proxies that best trace the most recent star formation, sSFR and $\agel$, are also the ones that show the greatest correlation with $\lamre$.  Once we control for these the correlation between $\lamre$ and $\lsfnu$ is not significant.  In contrast, there remains a weak but significant correlation between $\lamre$ and $\lsfnu$ if we control for age using $\agem$.  This is suggestive of $\lamre$ being most strongly related to the timescale of star formation shut-down.  We will explore these ideas in more detail below.

Note that above we look at correlations assuming linear $\lamre$.  \citet{2022ApJ...937..117F} argue that $\lamre$ is log-normally distributed (largely based on theoretical arguments regarding the spin of dark matter halos).  If we look for correlations between $\log(\lamre)$ and the other parameters, the qualitative trends are unchanged.  The correlation between $\log(\lamre)$ and age is still strongest ($r=-0.416$) and the correlation with $\lsfnu$ is not significant.  There is a residual correlation between $\log(\lamre)$ and $\lmstarnu$ that is significant ($r=-0.087$), but still substantially weaker than for age.

\subsection{Trends across the age-$\lamre$ plane}\label{sec:full_sample}

\begin{table*}
	\centering
	\caption{Results of K-S test comparisons between $\Delta\lamre$ distributions in different $\lsfnu$ intervals.  Both the K-S $D$ statistic and the probability of rejecting the null hypothesis that the distributions are the same, $P(<D)$, are given, along with the number of galaxies in each sample, $N_{\rm g}$.  Results for $\agel$, $\agem$ and $\lssfrnu$ are given.  We also calculate the results for two different minimum masses [column labeled $\log(M_{\rm *,min})$], of $\lmstar=9.5$ and 11.  Note that while tests against the full sample (labelled 'All') are included, the sub-samples are not statistically independent from the full sample.  Cases where a statistically significant difference ($P(<D)<0.01$) between the samples have their $P(<D)$ values highlighted in bold.}
	\label{tab:ks_sigma5}
	\begin{tabular}{ccrrrrrrrrrr} 
\hline
$\lsf$ & & & & \multicolumn{2}{c}{$<-0.5$} &\multicolumn{2}{c}{$-0.5$ to $ 0.5$} &\multicolumn{2}{c}{$ 0.5$ to $ 1.5$} &\multicolumn{2}{c}{$>1.5$}\\
\hline
 & Age & $\log(M_{\rm *,min})$ & $N_{\rm g}$  & $D$ & $P(<D)$ & $D$ & $P(<D)$  & $D$ & $P(<D)$  & $D$ & $P(<D)$\\
 \hline
All & $\agel$ &  9.5 & 1571& 0.087 &2.49e-01& 0.029 &8.85e-01& 0.040 &4.12e-01& 0.061 &4.26e-01\\
$-2.0$ to $-0.5$ & $\agel$ &  9.5 &  144& $-$ & $-$& 0.090 &3.00e-01& 0.115 &7.97e-02& 0.078 &6.20e-01\\
$-0.5$ to $ 0.5$ & $\agel$ &  9.5 &  520& $-$ & $-$& $-$ & $-$& 0.055 &3.20e-01& 0.062 &5.36e-01\\
$ 0.5$ to $ 1.5$ & $\agel$ &  9.5 &  675& $-$ & $-$& $-$ & $-$& $-$ & $-$& 0.098 &6.54e-02\\
$ 1.5$ to $ 3.0$ & $\agel$ &  9.5 &  232& $-$ & $-$& $-$ & $-$& $-$ & $-$& $-$ & $-$\\
\hline
 All & $\agel$ & 11.0 &  164& 0.488 &5.67e-01& 0.177 &1.84e-01& 0.115 &4.92e-01& 0.084 &9.39e-01\\
$-2.0$ to $-0.5$ & $\agel$ & 11.0 &    2& $-$ & $-$& 0.500 &5.71e-01& 0.500 &5.48e-01& 0.478 &6.45e-01\\
$-0.5$ to $ 0.5$ & $\agel$ & 11.0 &   46& $-$ & $-$& $-$ & $-$& 0.286 &1.67e-02& 0.217 &2.29e-01\\
$ 0.5$ to $ 1.5$ & $\agel$ & 11.0 &   70& $-$ & $-$& $-$ & $-$& $-$ & $-$& 0.119 &7.75e-01\\
$ 1.5$ to $ 3.0$ & $\agel$ & 11.0 &   46& $-$ & $-$& $-$ & $-$& $-$ & $-$& $-$ & $-$\\
\hline
 All & $\agem$ &  9.5 & 1571& 0.076 &4.10e-01& 0.068 &4.86e-02& 0.013 &1.00e+00& 0.152 &{ \bf 1.52e-04}\\
$-2.0$ to $-0.5$ & $\agem$ &  9.5 &  144& $-$ & $-$& 0.061 &7.72e-01& 0.078 &4.43e-01& 0.189 & {\bf 2.90e-03}\\
$-0.5$ to $ 0.5$ & $\agem$ &  9.5 &  520& $-$ & $-$& $-$ & $-$& 0.077 &5.88e-02& 0.214 & {\bf 6.70e-07}\\
$ 0.5$ to $ 1.5$ & $\agem$ &  9.5 &  675& $-$ & $-$& $-$ & $-$& $-$ & $-$& 0.160 &{\bf 2.37e-04}\\
$ 1.5$ to $ 3.0$ & $\agem$ &  9.5 &  232& $-$ & $-$& $-$ & $-$& $-$ & $-$& $-$ & $-$\\
\hline
All & $\agem$ & 11.0 &  164& 0.482 &5.89e-01& 0.134 &4.89e-01& 0.041 &1.00e+00& 0.160 &2.84e-01\\
$-2.0$ to $-0.5$ & $\agem$ & 11.0 &    2& $-$ & $-$& 0.500 &5.71e-01& 0.500 &5.48e-01& 0.500 &5.71e-01\\
$-0.5$ to $ 0.5$ & $\agem$ & 11.0 &   46& $-$ & $-$& $-$ & $-$& 0.158 &4.33e-01& 0.283 &5.03e-02\\
$ 0.5$ to $ 1.5$ & $\agem$ & 11.0 &   70& $-$ & $-$& $-$ & $-$& $-$ & $-$& 0.184 &2.64e-01\\
$ 1.5$ to $ 3.0$ & $\agem$ & 11.0 &   46& $-$ & $-$& $-$ & $-$& $-$ & $-$& $-$ & $-$\\
\hline
All & $sSFR$ &  9.5 & 1596& 0.113 &5.26e-02& 0.069 &4.07e-02& 0.076 &{\bf 7.28e-03}& 0.045 &7.82e-01\\
$-2.0$ to $-0.5$ & $sSFR$ &  9.5 &  153& $-$ & $-$& 0.064 &6.90e-01& 0.178 &{\bf 6.26e-04}& 0.108 &2.21e-01\\
$-0.5$ to $ 0.5$ & $sSFR$ &  9.5 &  530& $-$ & $-$& $-$ & $-$& 0.144 &{\bf 7.52e-06}& 0.065 &4.84e-01\\
$ 0.5$ to $ 1.5$ & $sSFR$ &  9.5 &  684& $-$ & $-$& $-$ & $-$& $-$ & $-$& 0.114 &2.13e-02\\
$ 1.5$ to $ 3.0$ & $sSFR$ &  9.5 &  229& $-$ & $-$& $-$ & $-$& $-$ & $-$& $-$ & $-$\\
\hline
All & $sSFR$ & 11.0 &  173& 0.630 &2.82e-01& 0.077 &9.57e-01& 0.082 &8.33e-01& 0.105 &7.68e-01\\
$-2.0$ to $-0.5$ & $sSFR$ & 11.0 &    2& $-$ & $-$& 0.620 &3.17e-01& 0.707 &1.89e-01& 0.543 &4.49e-01\\
$-0.5$ to $ 0.5$ & $sSFR$ & 11.0 &   50& $-$ & $-$& $-$ & $-$& 0.133 &6.28e-01& 0.156 &5.40e-01\\
$ 0.5$ to $ 1.5$ & $sSFR$ & 11.0 &   75& $-$ & $-$& $-$ & $-$& $-$ & $-$& 0.181 &2.63e-01\\
$ 1.5$ to $ 3.0$ & $sSFR$ & 11.0 &   46& $-$ & $-$& $-$ & $-$& $-$ & $-$& $-$ & $-$\\
\hline
	\end{tabular}
\end{table*}

Given that age (or sSFR) correlates most strongly with $\lamre$, in Fig. \ref{fig:lam_age_sigma5} we show the distribution of SAMI galaxies in the $\lamre$ vs.\ age plane.  For age we show both light-weighted ($\agel$; panels a, d, and g) and mass-weighted ($\agem$; panels b, e and h) from full spectral fitting, as well as sSFR (panels c, f and i).  The results from using Lick-index based SSP ages \citep{2017MNRAS.472.2833S} are not shown, but are qualitatively similar to the light-weighted age estimates.  In each row in Fig.\ \ref{fig:lam_age_sigma5} we show one of our different environmental metrics.  As seen above [and previously pointed out by \citet{2018NatAs...2..483V}], there is a strong relation between age and $\lamre$, such that galaxies with older ages [or lower sSFR - see also \citet{2021MNRAS.503.4992F}] have lower spin.  This is true for both light-weighted and mass-weighed ages.  Binning the data in age, we calculate the median spin (large-black points in Fig.\ \ref{fig:lam_age_sigma5}).  The median trend of $\lamre$ vs.\ age is well approximated by a sigmoid function of the form given by Eq.\ \ref{eq:sigmoid}.

The fitted parameters for our full sample for Eq.\ \ref{eq:sigmoid} are given in Table \ref{tab:sigmoid_fit}. For the three different age proxies the asymptotic value at young ages ($\lambda_0$) is well constrained and reasonably consistent.  There is greater variation on the asymptotic value at old age ($\lambda_1)$, as the trend does not as clearly flatten at old age.  If we remove slow rotators, then the overall trend is largely unchanged, although the $\lambda_1$ value becomes slightly higher.  The scatter around the median relation is relatively small, particularly noting that we are analysing observed $\lamre$ values that have the impact random inclination imprinted on them as well as measurement uncertainty and beam-smearing correction.  The RMS scatter is $\simeq0.15$ for $\agel$ and $sSFR$, and a slightly larger (0.17) for $\agem$.  This is significantly higher than the median measurement uncertainty on $\lamre$, incorporating the beam-smearing correction, which is 0.032.

In the panels in Fig.\ \ref{fig:lam_age_sigma5} we colour the points using different environmental metrics.  In panels a, b and c we use $\lsfnu$, in panels d, e and f we use $\log(\mhalo)$ and in panels g, h and i we use environmental class.  The large coloured points in each case are the median $\lamre$ values in bins of age and environment, with error bars giving the error on the median.  The broad conclusion from the comparison of spin, age and environment is consistent with the correlation analysis above.  Once age is taken into account there is relatively little remaining dependence on environment, although there are some residual environmental trends that we will discuss below.

We split the galaxies into four different environmental intervals with $\lsf$ values $<-0.5$, $-0.5$ to 0.5, 0.5 to 1.5 and $>1.5$ (Figs.\ \ref{fig:lam_age_sigma5}a, b and c).  For the light-weighted age and sSFR cases (Figs.\ \ref{fig:lam_age_sigma5}a and c), there is no indication that galaxies in different $\lsfnu$ intervals follow different paths in the $\lamre$-age plane.  We perform a K-S test comparing the distributions of $\Delta\lamre$ (the residual after removing the median $\lamre$-age trend) for different environments.  The results of this test are listed in Table \ref{tab:ks_sigma5}.  For the light-weighted ages none of the K-S tests show significant differences between the environmental samples (i.e.\ all P-values are greater than 0.01), confirming the visual impression from Fig.\ \ref{fig:lam_age_sigma5}a.  When using sSFR the galaxies at $0.5<\lsf<1.5$ (green points in Fig.\ \ref{fig:lam_age_sigma5}c) do show a small offset such that they have slightly higher $\lamre$ than other samples.  This difference is not very obvious in Fig.\ \ref{fig:lam_age_sigma5}c, but is seen as a significant K-S test in Table \ref{tab:ks_sigma5}.   

When we consider the mass-weighted ages (Fig.\ \ref{fig:lam_age_sigma5}b) we start to see a difference at ages $\agem>8$\,Gyr.  Galaxies in denser environments have slightly lower spin.  This is borne out by a significant difference found using the K-S tests for $\Delta\lamre$ (far right column in Table \ref{tab:ks_sigma5}).  The differences between the galaxies in the highest density [$\lsf>1.5$] and the others is always significant, with P-values of $2.9e10^{-3}$ to $6.7e10^{-7}$.

\begin{table}
	\centering
	\caption{Results of K-S test comparisons between $\Delta\lamre$ distributions in different $\lmhalo$ intervals.  Both the K-S $D$ statistic and the probability of rejecting the null hypothesis that the distributions are the same, $P(<D)$, are given, along with the number of galaxies in each sample, $N$.  Subscripts 1 and 2 represent the two samples compared, with $M_{h1}$ and $M_{h2}$ columns containing the range in $\lmhalo$ for the two samples considered in each row.  Both light weighted ($\agel$) and mass-weighted ($\agem$) are given.  Cases where there is a statistically significant difference ($P(<D)<0.01$) between the samples have their $P(<D)$ values highlighted in bold.}
	\label{tab:ks_mhalo}
	\begin{tabular}{cccrrrrr} 
\hline
 ${\mhalo}_1$ & ${\mhalo}_2$ & Age & $N_1$ & $N_2$ & $M_{*,min}$ & $D$ & $P(<D)$\\ 
 \hline
 0--13 & 13--14 & $\agel$ & 256 & 315 &  9.5 & 0.094 & 1.50e-01\\
 0--13 & 14--16 & $\agel$ & 256 & 699 &  9.5 & 0.096 & 5.87e-02\\
13--14 & 14--16 & $\agel$ & 315 & 699 &  9.5 & 0.098 & 2.96e-02\\
\hline
 0--13 & 13--14 & $\agel$ &  19 &  69 & 11.0 & 0.140 & 8.86e-01\\
 0--13 & 14--16 & $\agel$ &  19 &  71 & 11.0 & 0.234 & 3.22e-01\\
13--14 & 14--16 & $\agel$ &  69 &  71 & 11.0 & 0.191 & 1.32e-01\\
\hline
 0--13 & 13--14 & $\agem$ & 256 & 315 &  9.5 & 0.077 & 3.57e-01\\
 0--13 & 14--16 & $\agem$ & 256 & 699 &  9.5 & 0.092 & 7.77e-02\\
13--14 & 14--16 & $\agem$ & 315 & 699 &  9.5 & 0.089 & 5.88e-02\\
\hline
 0--13 & 13--14 & $\agem$ &  19 &  69 & 11.0 & 0.210 & 4.56e-01\\
 0--13 & 14--16 & $\agem$ &  19 &  71 & 11.0 & 0.135 & 9.07e-01\\
13--14 & 14--16 & $\agem$ &  69 &  71 & 11.0 & 0.190 & 1.35e-01\\
\hline
 0--13 & 13--14 & $sSFR$ & 261 & 328 &  9.5 & 0.068 & 4.90e-01\\
 0--13 & 14--16 & $sSFR$ & 261 & 698 &  9.5 & 0.141 & {\bf 8.73e-04}\\
13--14 & 14--16 & $sSFR$ & 328 & 698 &  9.5 & 0.143 & {\bf 1.79e-04}\\
\hline
 0--13 & 13--14 & $sSFR$ &  20 &  77 & 11.0 & 0.264 & 1.78e-01\\
 0--13 & 14--16 & $sSFR$ &  20 &  72 & 11.0 & 0.153 & 8.03e-01\\
13--14 & 14--16 & $sSFR$ &  77 &  72 & 11.0 & 0.248 & 1.63e-02\\
\hline
\end{tabular}
\end{table}

Turning to alternative environmental metrics, in Figs.\ \ref{fig:lam_age_sigma5}d, e and f we show the $\lamre$-age plane, but this time colour-coded by $\lmhalonu$.  When separating by halo mass, there are fewer galaxies included, as we exclude those that are not part of a group defined within GAMA, or not within the SAMI clusters.  The large coloured points in \ref{fig:lam_age_sigma5}d, e and f show the median values in 3 halo mass intervals, $\lmhalo<13$ (blue), $13<\lmhalo<14$ (cyan) and $\lmhalo>14$ (red).  Once again, the dominant trend is between $\lamre$ and age, with little obvious dependency on halo mass.  The exception is when using sSFR as our age proxy. In this case galaxies with $\lssfr<-12$ are somewhat separated depending on halo mass.  In particular, galaxies in high-mass halos ($\lmhalo>14$) have higher spin than those in lower mass halos [$P(<D)=$1.8e-4 and 8.7e-4 in Table \ref{tab:ks_mhalo}].  It is worth noting that $\lssfr$ values below --12 are not physically meaningful measurements of star formation rate and these galaxies can all be considered to be fully passive.

\begin{table}
	\centering
	\caption{Results of K-S test comparisons between $\Delta\lamre$ distributions in different classes (isolated, central or satellite).  Both the K-S $D$ statistic and the probability of rejecting the null hypothesis that the distributions are the same, $P(<D)$, are given, along with the number of galaxies in each sample, $N$.  Subscripts 1 and 2 represent the two samples compared.  Both light weighted ($\agel$) and mass-weighted ($\agem$) are given.  Cases where a statistically significant difference ($P(<D)<0.01$) between the samples have their $P(<D)$ values highlighted in bold.}
	\label{tab:ks_satcent}
	\begin{tabular}{cccrrrrr} 
\hline
 Class$_1$ & Class$_2$ & Age & $N_1$ & $N_2$ & $M_{*,min}$ & $D$ & $P(<D)$\\ 
 \hline
iso & cent & $\agel$ & 318 & 399 &  9.5 & 0.140 & {\bf 1.62e-03}\\
iso & sat & $\agel$ & 318 & 887 &  9.5 & 0.073 & 1.54e-01\\
cent & sat & $\agel$ & 399 & 887 &  9.5 & 0.162 & {\bf 9.21e-07}\\
\hline
iso & cent & $\agel$ &  12 & 102 & 11.0 & 0.294 & 2.61e-01\\
iso & sat & $\agel$ &  12 &  58 & 11.0 & 0.365 & 1.08e-01\\
cent & sat & $\agel$ & 102 &  58 & 11.0 & 0.272 & {\bf 6.47e-03}\\
\hline
iso & cent & $\agem$ & 318 & 399 &  9.5 & 0.165 & {\bf 1.10e-04}\\
iso & sat & $\agem$ & 318 & 887 &  9.5 & 0.107 & {\bf 8.19e-03}\\
cent & sat & $\agem$ & 399 & 887 &  9.5 & 0.145 & {\bf 1.57e-05}\\
\hline
iso & cent & $\agem$ &  12 & 102 & 11.0 & 0.333 & 1.48e-01\\
iso & sat & $\agem$ &  12 &  58 & 11.0 & 0.198 & 7.52e-01\\
cent & sat & $\agem$ & 102 &  58 & 11.0 & 0.324 & {\bf 5.90e-04}\\
\hline
iso & cent & $sSFR$ & 330 & 415 &  9.5 & 0.176 & {\bf 1.92e-05}\\
iso & sat & $sSFR$ & 330 & 888 &  9.5 & 0.145 & {\bf 6.51e-05}\\
cent & sat & $sSFR$ & 415 & 888 &  9.5 & 0.199 & {\bf 2.83e-10}\\
\hline
iso & cent & $sSFR$ &  14 & 110 & 11.0 & 0.174 & 7.84e-01\\
iso & sat & $sSFR$ &  14 &  60 & 11.0 & 0.424 & 2.34e-02\\
cent & sat & $sSFR$ & 110 &  60 & 11.0 & 0.359 & {\bf 5.79e-05}\\
\hline
    \end{tabular}
\end{table}

\begin{figure*}
	\includegraphics[trim=0 5mm 0 0,clip,width=16cm]{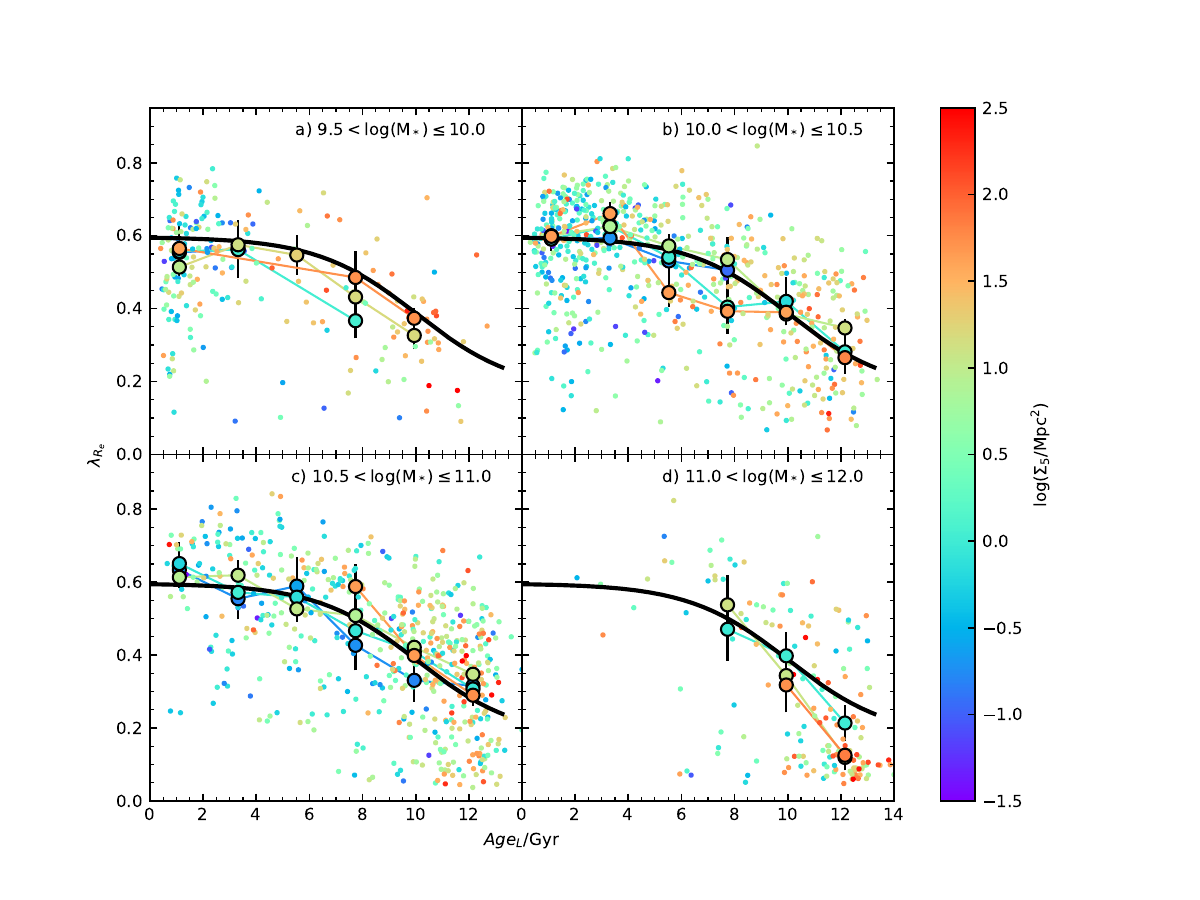}
    \caption{The relation between $\lamre$ and light-weighted age, $\agel$, for SAMI galaxies, colour coded by environment using $\lsfnu$, sub-divided by stellar mass.  The sigmoid function fit to the total population is shown by the thick black line.  The large coloured points are the median $\lamre$ in age bins, but separated into different intervals in environment, using the same colour coding as the smaller points.  Error bars show the uncertainty on the median, calculated using the 68th percentile width of the distribution.}
    \label{fig:lam_agel_mass_sigma5}
\end{figure*}

Our third environmental metric examined is splitting the galaxies by {\it environment class}.  In Fig.\ \ref{fig:lam_age_sigma5}g, h and i we show the spin-age plane, with the points colour coded by class.  When separated by class, we start to see more significant differences between galaxy populations.  The overall trend in age still dominates, but galaxies of different classes are more separated and this is most clear at old ages (for $\agel$, $\agem$ and $\lssfrnu$).  At old ages ($\gtrsim8$\,Gyr), central galaxies (orange points) have the lowest $\lamre$, while isolated galaxies (green points) have the highest $\lamre$.  Satellites (blue points) typically sit in between centrals and isolated galaxies (although for $sSFR$ isolated and satellite galaxies appear similar).  At ages below $\sim8$\,Gyr there is no obvious difference between the three classes.  As before, we examine the difference between classes statistically using the K-S test on the distributions of $\Delta\lamre$ (see Table \ref{tab:ks_satcent}). In contrast to the other environmental metrics, we find significant differences in many cases.  In particular, centrals are usually significantly different from satellites or isolated galaxies.  However, the difference between satellites and isolated galaxies is only sometimes significant.

If we remove slow rotators from the analysis we find qualitatively similar results regarding halo mass, with no significant trends once age is accounted for.  For environmental class, removing slow rotators reduces the significance of the difference between centrals and the other galaxies.  This is not surprising given that we know that a high fraction of central galaxies are slow rotators \citep[e.g.][]{2021MNRAS.508.2307V}.

In summary, we find that across a range of different metrics the change in $\lamre$ due to environment is small compared to the trends between $\lamre$ and age.  The only convincing cases where we find a residual environmental trend are i) comparing central, satellite and isolated galaxies for all age metrics; and ii) comparing different halo masses when using sSFR as our age metric.  It could be argued that difference between isolated and central galaxies is surprising, given that isolated galaxies are, in fact, centrals, where the the satellites are too faint to be detected in the environment defining population.  However, the halo mass of isolated galaxies will typically be lower than that of centrals.  The isolated galaxies are also likely to have lower stellar mass than the centrals and so far we have not separated the population by galaxy stellar mass.  That said, the recent work by \citet{2021MNRAS.508.2307V} shows that the slow rotator fraction is lower for isolated galaxies than centrals, even when controlling for stellar mass.  It is intriguing that age appears more important than environment.  To explore this further, we will next look at $\lamre$-age relations in narrow mass intervals, to see if particular mass ranges dominate any visible trends.

\begin{figure*}
	\includegraphics[trim=0 5mm 0 0,clip,width=16cm]{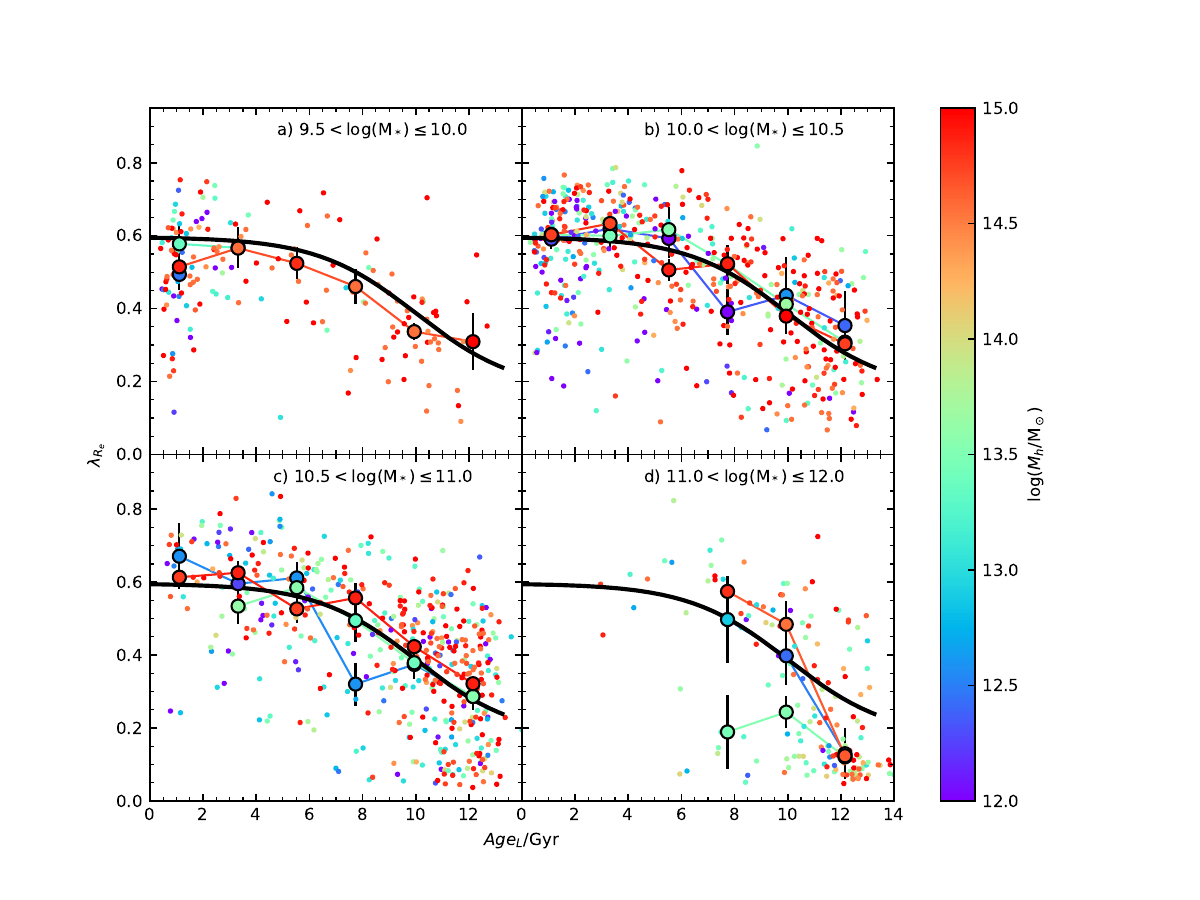}
    \caption{The relation between $\lamre$ and light-weighted age, $\agel$, for SAMI galaxies, colour coded by halo mass, $\lmhalonu$, sub-divided by stellar mass.  The sigmoid function fit to the total population is shown by the thick black line.  The large coloured points are the median $\lamre$ in age bins, but separated by halo mass, using the same colour coding as the smaller points.  Error bars show the uncertainty on the median, calculated using the 68th percentile width of the distribution.}
    \label{fig:lam_agel_mass_halo}
\end{figure*}

\begin{figure*}
	\includegraphics[trim=0 5mm 0 0,clip,width=16cm]{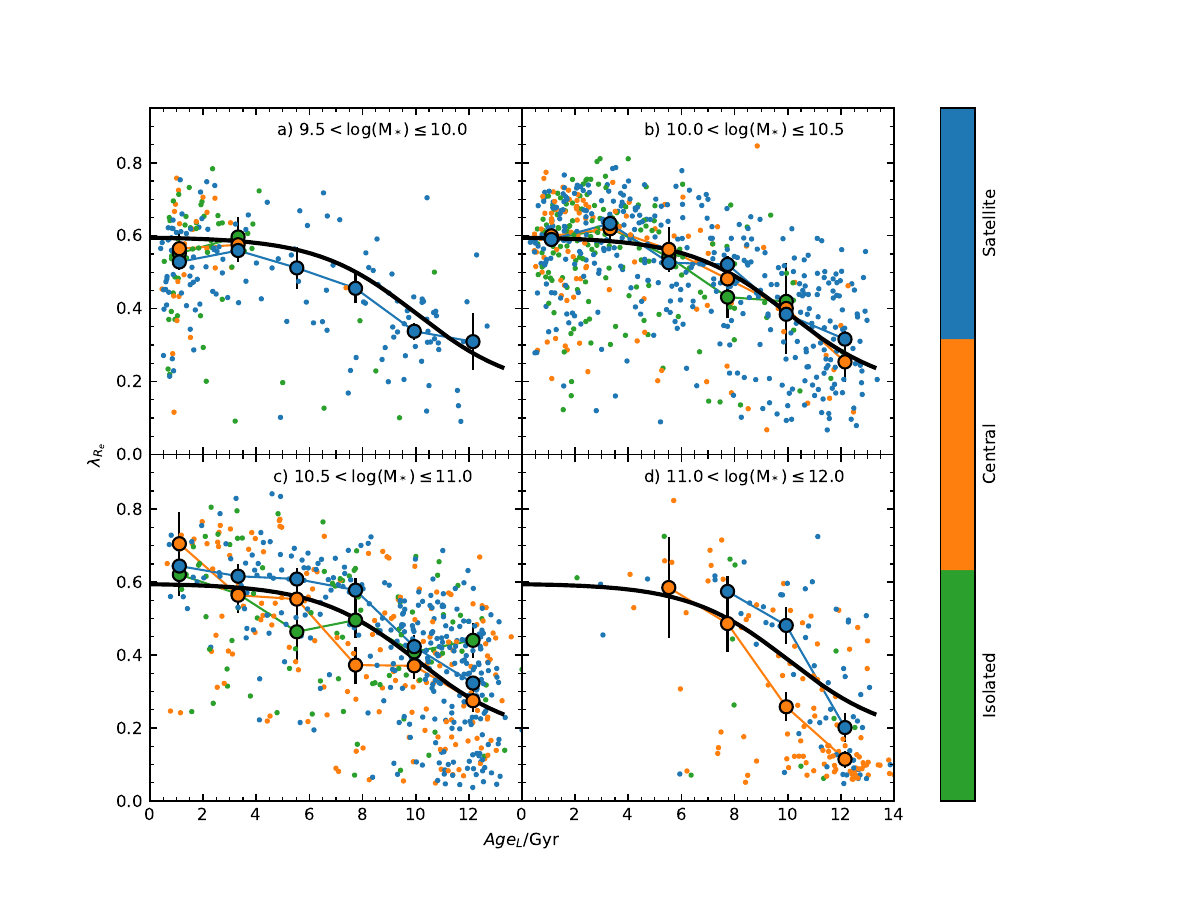}
    \caption{The relation between $\lamre$ and light-weighted age, $\agel$, for SAMI galaxies, colour coded by environmental class, sub-divided by stellar mass.  The sigmoid function fit to the total population is shown by the thick black line.  The large coloured points are the median $\lamre$ in age bins, but separated by environmental class, using the same colour coding as the smaller points.  Error bars show the uncertainty on the median, calculated using the 68th percentile width of the distribution.}
    \label{fig:lam_agel_mass_satcent}
\end{figure*}

\subsection{Trends as a function of mass}
\label{sec:mass}

Now we subdivide the sample by stellar mass, using four intervals in $\lmstar$: $9.5-10.0$, $10.0-10.5$, $10.5-11.0$ and $11.0-12.0$.  We do this to examine whether there may be trends that are obscured when considering the sample as a whole.  In particular, at high mass ($\lmstar\gtrsim11$) the slow rotator fraction increases significantly, so the behaviour in this regime may be different to the rest of the population.  We will only show the results for $\agel$ in this section.  We make this choice as i) $\agel$ or $sSFR$ correlate better with $\lamre$ than $\agem$; ii) $\agel$ is well defined for almost all galaxies, while $sSFR$ is not well defined below $\lssfr\sim-12$; and iii) results from $\agel$ and $sSFR$ are qualitatively similar.

The trends subdivided by mass using $\agel$ and $\lsfnu$ are shown in Fig. \ref{fig:lam_agel_mass_sigma5}.  The sample is also subdivided into four different environmental intervals with $\lsf$ values $<-0.5$, $-0.5$ to 0.5, 0.5 to 1.5 and $>1.5$ (the same as used in Section \ref{sec:full_sample}). In mass intervals below $\lmstar=11$ galaxies in different environments all broadly follow the same relation between $\agel$ and $\lamre$ (the thick black line shows the median relation for the full sample so is the same for each panel).  Galaxies at low mass (Fig.\ \ref{fig:lam_agel_mass_sigma5}a) lie slightly below the median relation.  This is due to the known weak relation between $\lamre$ and mass \citep[e.g.][]{2021MNRAS.508.2307V} at low mass.  In the highest mass interval (Fig.\ \ref{fig:lam_agel_mass_sigma5}d, $11.0<\lmstar\leq12.0$) there is an increased fraction of galaxies below the median relation, particularly at old ages, $\agel>10$\,Gyr.  These galaxies tend to have low $\lamre$, below $\sim0.2$, and would mostly be classified as slow rotators using standard diagnostics \citep[e.g.][]{2016ARA&A..54..597C}.  However, there is no obvious environmental trend once the age trend is accounted for.  K-S tests on $\Delta\lamre$ do not find a significant difference when limited by mass (e.g. a minimum mass of $\lmstar=11.0$ as seen in Table \ref{tab:ks_sigma5}). 

When we consider the trends in $\lamre$ vs.\ $\agem$, subdivided by mass (not shown), we find that the residual trend with $\Sigma_5$ is mostly visible at intermediate masses, $\lmstar=10-10.5$ and $10.5-11.0$.  This result suggests the residual trends for $\agem$ are not purely driven by high mass slow rotators.

When we instead use halo mass as our environmental metric, but split by galaxy stellar mass (Fig.\ \ref{fig:lam_agel_mass_halo}) we only see evidence of an environmental difference in the highest stellar mass interval ($\lmstar>11$).  Overall, the K-S test for $\Delta\lamre$ does not show significant differences (Table \ref{tab:ks_mhalo}), but looking in detail (Fig.\ \ref{fig:lam_agel_mass_halo}d) we see that for the highest stellar mass interval the median spin of the oldest galaxies $\agel\sim12-14$\,Gyr is low ($\lamre\sim0.1$) for all halo masses.  However, for younger ages ($\sim7-11$\,Gyr) it is the galaxies in intermediate mass halos (cyan points in Figs.\ \ref{fig:lam_agel_mass_halo}d) that have the lowest $\lamre$.  This implies that if environment does drive the spin of massive galaxies, having the lowest spin galaxies in the highest mass halos may not be a sufficient explanation.

In Fig.\ \ref{fig:lam_agel_mass_satcent} we show the SAMI galaxies subdivided by mass in the $\lamre$ vs. $\agel$ plane, but using environmental class (isolated, central or satellite) as the environmental indicator.  When analysing the full sample (Section \ref{sec:full_sample}) environmental class was the metric that showed the most significant differences (see Table \ref{tab:ks_satcent}), with central galaxies having the lowest $\lamre$.  When subdividing by mass we see that this trend is clear in the highest mass interval (e.g. Fig.\ \ref{fig:lam_agel_mass_satcent}d) and also appears present the second-highest mass interval (Fig.\ \ref{fig:lam_agel_mass_satcent}c).    However, the trend is not visible at masses below $\lmstar=10.5$.    The $\Delta\lamre$ analysis (Table \ref{tab:ks_satcent}) shows that the difference between centrals and satellites continues to be significant, even when restricting the sample to $\lmstar>11$ (for $\agel$, $\agem$ and $sSFR$).  Interestingly, if we use a limit of $\lmstar>10.5$ (not listed in Table \ref{tab:ks_satcent}) we find the highest significance when comparing centrals and satellites with $D=0.224$ and $P(<D)=3.3e-07$.  

If we compare the $\lamre$-age distribution of $\lmstar>11$ galaxies when flagged by halo mass (Fig.\ \ref{fig:lam_agel_mass_halo}d) to the case when flagged by environmental class (Fig.\ \ref{fig:lam_agel_mass_satcent}d) then we see that the population of low $\lamre$ galaxies at intermediate halo masses with ages $\sim7-11$\,Gyr are almost all central galaxies.  It is unsurprising that these massive galaxies are almost all centrals given that they live in halos in the mass range $\lmhalo=13-14$ \citep[due to the stellar-to-halo-mass relation; e.g.][]{2013ApJ...778...93T}.  At face value this leads to two potential causes.  The first is that the intermediate mass halos ($\lmhalo=13-14$) are optimal locations to form galaxies with low $\lamre$, e.g.\ because they are efficient locations for merging.  The alternative is that simply being a high-mass central galaxy is sufficient to increase the likelihood of having low $\lamre$.    We will discuss these points further below.

\section{Comparisons to simulations}\label{sec:sims}

\begin{figure*}
	\includegraphics[trim=10mm 5mm 5mm 5mm, clip, width=16cm]{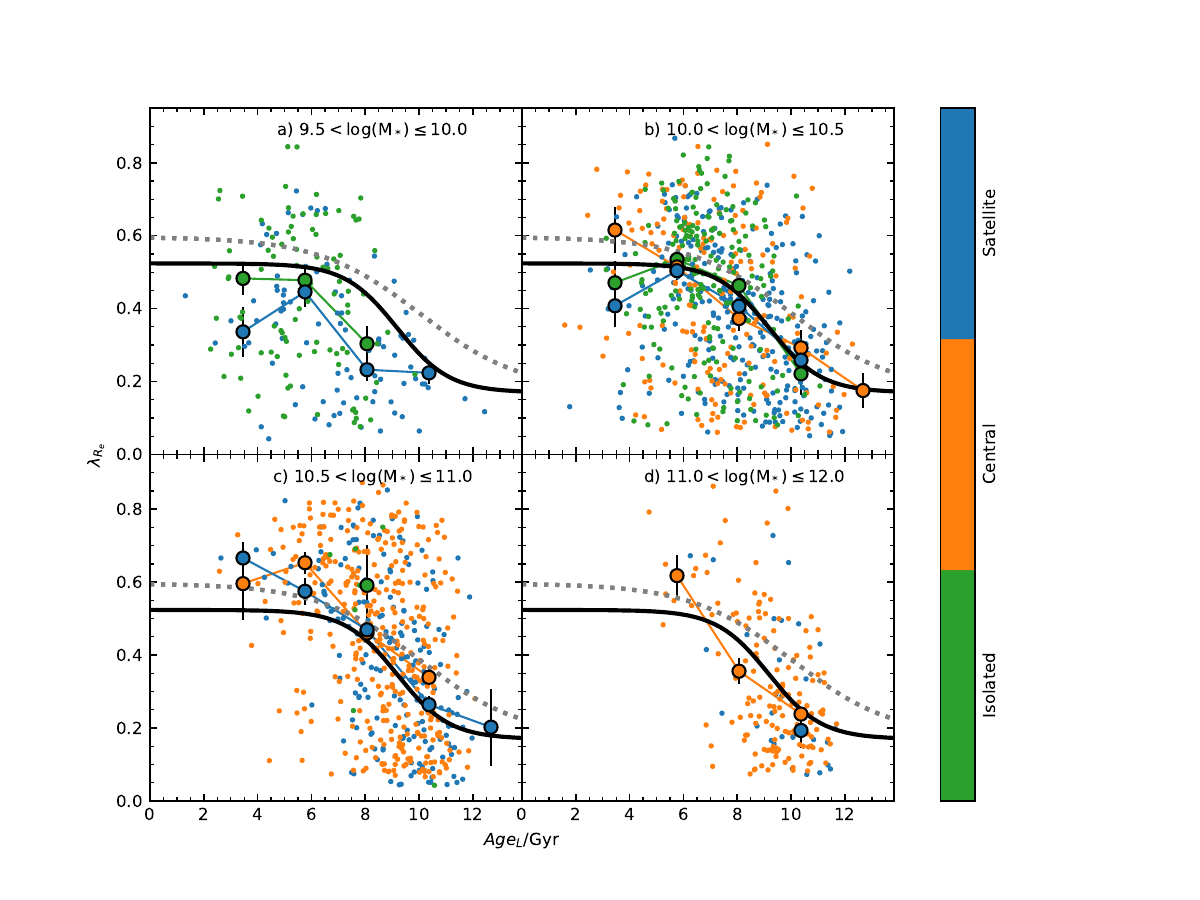}
    \caption{The relation between $\lamre$ and age for EAGLE simulated galaxies, colour coded by environment using satellite vs. central class and sub-divided by mass.  The thick black line shows the best fit sigmoid function to the median spin in age bins (for all galaxies).  The large coloured points are the median spin in age bins, but separated into different intervals in environment, using the same colour coding as the smaller points.  Error bars show the uncertainty on the median, calculated using the 68th percentile width of the distribution.  The grey dotted line is the best fit $\lamre$ vs.\ $\agel$ relation for SAMI.}
    \label{fig:lam_age_satcent_eagle}
\end{figure*}

\begin{table}
	\centering
	\caption{The results of our correlation analysis between 4 parameters: $\lmstarnu$, $\lsfnu$, $\lamre$ and an age proxy for the EAGLE simulation.  The age proxy is either light-weighted age ($\agel$), mass-weighted age ($\agem$) or $\lssfrnu$.  For each pair of parameters $(A,B)$ we list the results of a full correlation analysis (just correlating the two parameters) and a partial correlation analysis (accounting for the correlations between other variables).  Correlations that are {\it not} statistically significant ($P$-value$>0.01$) are highlighted in bold.  We list results for both the $r$-band and stellar-mass weighted $\lamre$, mass matched to SAMI above $\lmstar=10$.}
	\label{tab:corr_four_full_eagle}
	\begin{tabular}{ccrrrr} 
\hline
& & \multicolumn{2}{c}{full correlation} &\multicolumn{2}{c}{partial correlation}\\
A & B & $r$ & $P$-value & $r$ & $P$-value \\
\hline
\multicolumn{4}{l}{EAGLE $r$-band weighted $\lamre$:}\\
\hline
$\lmstarnu$ & $\agel$ &  0.400 & 2.76e-55 &   0.392 & 7.78e-53 \\
$\lmstarnu$ & $\lsfnu$ &  0.098 & 2.17e-04 &   0.002 & {\bf 9.49e-01} \\
$\lmstarnu$ & $\lamre$ & -0.071 & 7.92e-03 &   0.074 & 5.56e-03 \\
$\agel$ & $\lsfnu$ &  0.250 & 1.83e-21 &   0.204 & 1.22e-14 \\
$\agel$ & $\lamre$ & -0.336 & 1.36e-38 &  -0.320 & 6.49e-35 \\
$\lsfnu$ & $\lamre$ & -0.125 & 2.33e-06 &  -0.045 & {\bf 8.79e-02} \\
\hline
$\lmstarnu$ & $\agem$ &  0.420 & 2.20e-61 &   0.407 & 2.28e-57 \\
$\lmstarnu$ & $\lsfnu$ &  0.098 & 2.17e-04 &   0.015 & {\bf 5.69e-01} \\
$\lmstarnu$ & $\lamre$ & -0.071 & 7.92e-03 &   0.014 & {\bf 6.09e-01} \\
$\agem$ & $\lsfnu$ &  0.204 & 9.21e-15 &   0.163 & 8.17e-10 \\
$\agem$ & $\lamre$ & -0.194 & 1.78e-13 &  -0.164 & 5.60e-10 \\
$\lsfnu$ & $\lamre$ & -0.125 & 2.33e-06 &  -0.089 & 7.86e-04 \\
\hline
$\lmstarnu$ & $\lssfrnu$ & -0.230 & 2.37e-18 &  -0.211 & 1.11e-15 \\
$\lmstarnu$ & $\lsfnu$ &  0.098 & 2.17e-04 &   0.005 & {\bf 8.39e-01} \\
$\lmstarnu$ & $\lamre$ & -0.071 & 7.92e-03 &   0.067 & {\bf 1.18e-02} \\
$\lssfrnu$ & $\lsfnu$ & -0.379 & 2.96e-49 &  -0.364 & 2.62e-45 \\
$\lssfrnu$ & $\lamre$ &  0.549 & 1.26e-111 &   0.546 & 2.07e-110 \\
$\lsfnu$ & $\lamre$ & -0.125 & 2.33e-06 &   0.106 & 7.06e-05 \\
\hline
\multicolumn{4}{l}{EAGLE mass weighted $\lamre$:}\\
\hline
$\lmstarnu$ & $\agel$ &  0.400 & 2.76e-55 &   0.392 & 5.61e-53 \\
$\lmstarnu$ & $\lsfnu$ &  0.098 & 2.17e-04 &  -0.004 & {\bf 8.79e-01} \\
$\lmstarnu$ & $\lamre$ & -0.041 & {\bf 1.20e-01} &   0.063 & {\bf 1.82e-02} \\
$\agel$ & $\lsfnu$ &  0.250 & 1.83e-21 &   0.233 & 9.58e-19 \\
$\agel$ & $\lamre$ & -0.243 & 1.85e-20 &  -0.250 & 1.98e-21 \\
$\lsfnu$ & $\lamre$ & -0.025 & {\bf 3.45e-01} &   0.038 & {\bf 1.54e-01} \\
\hline
$\lmstarnu$ & $\agem$ &  0.420 & 2.20e-61 &   0.410 & 4.02e-58 \\
$\lmstarnu$ & $\lsfnu$ &  0.098 & 2.17e-04 &   0.014 & {\bf 6.03e-01} \\
$\lmstarnu$ & $\lamre$ & -0.041 & {\bf 1.20e-01} &   0.026 & {\bf 3.27e-01} \\
$\agem$ & $\lsfnu$ &  0.204 & 9.21e-15 &   0.179 & 1.17e-11 \\
$\agem$ & $\lamre$ & -0.155 & 5.18e-09 &  -0.150 & 1.51e-08 \\
$\lsfnu$ & $\lamre$ & -0.025 & {\bf 3.45e-01} &   0.006 & {\bf 8.13e-01} \\
\hline
$\lmstarnu$ & $\lssfrnu$ & -0.230 & 2.37e-18 &  -0.209 & 2.27e-15 \\
$\lmstarnu$ & $\lsfnu$ &  0.098 & 2.17e-04 &  0.008 & {\bf 7.69e-01} \\
$\lmstarnu$ & $\lamre$ & -0.041 & {\bf 1.20e-01} &   0.039 & {\bf 1.41e-01} \\
$\lssfrnu$ & $\lsfnu$ & -0.379 & 2.96e-49 &  -0.383 & 1.63e-50 \\
$\lssfrnu$ & $\lamre$ &  0.342 & 7.32e-40 &   0.359 & 4.84e-44 \\
$\lsfnu$ & $\lamre$ & -0.025 & {\bf 3.45e-01} &   0.119 & 7.20e-06 \\
\hline
\end{tabular}
\end{table}

To further understand the physical implications of the SAMI results presented above, we carry out a similar analysis using simulated galaxies from EAGLE \citep{2015MNRAS.446..521S}.  Our first step will be to look at the simulation results at $z=0$ and see whether they show signatures consistent with the observational data.  Then we will examine how the trends evolve with redshift, in order to make predictions of the signal we would expect to see from observational data at earlier epochs.  We note that \citet{2019MNRAS.484..869V} provides a detailed comparison between SAMI galaxies and simulations that we do not aim to repeat here.

\subsection{Simulation comparisons at $z=0$}\label{sec:lowzsims}

We first carry out a correlation analysis, as was done in Section \ref{sec:correlation}.  Results of the full and partial correlations between $\lamre$, $\lmstarnu$, $\lsfnu$ and age (using $\agel$, $\agem$ and $sSFR$ as age proxies) are shown in Table \ref{tab:corr_four_full_eagle}.    The top half of Table \ref{tab:corr_four_full_eagle} shows the results using the $r$-band weighted $\lamre$ from EAGLE, mass matched to SAMI above $\lmstar=10$.  The mass matching is done by randomly selecting one EAGLE galaxy for each SAMI galaxy, limiting the stellar mass difference to be less than 0.1 dex. 
The EAGLE data show the same general trends that we see with SAMI; that is, age (or sSFR) is the parameter that most strongly correlates with $\lamre$ for both the full and partial correlations.  In particular, the correlation between $\lamre$ and $\lsfnu$ is much weaker ($r=-0.045$) than the correlation with $\agel$ ($r=-0.320$).  The partial correlations between our age proxies and $\lmstarnu$ are also often not significant.  Like the SAMI data, EAGLE shows a stronger $\lamre$ correlation with $\agel$ and $\lssfrnu$ than $\agem$.

An advantage of the EAGLE simulations is that we can directly measure a mass-weighted $\lamre$ as well as the light-weighted version.  This allows us to explore whether our use of light-weighted $\lamre$ is what leads to the strongest correlation being with light-weighted stellar population age.  The mass-weighted $\lamre$ correlation results (bottom half of Table \ref{tab:corr_four_full_eagle}) show that the partial correlation of $\lamre$ is strongest for $sSFR$ ($r=0.350$), followed by $\agel$ ($r=-0.250$) and then $\agem$ ($r=-0.150$).  The correlations between mass-weighted $\lamre$ and age proxies are consistently stronger than the correlations between mass-weighted $\lamre$ and $\lmstarnu$ or $\lsfnu$.  Therefore, at least within the EAGLE simulated data, the use of light-weighted $\lamre$ is not the primary reason that $\agel$ and $sSFR$ appear to be the primary drivers of spin.  Similarly, as the EAGLE ages do not have measurement uncertainties (at least not in the same sense as the observational data), the fact that the correlation is always stronger for light weighted age is suggestive that the same trend in the observations is not associated with larger uncertainty on mass weighted ages (see discussion in Section \ref{sec:corrmasswt}).

Fig. \ref{fig:lam_age_satcent_eagle} shows the EAGLE galaxies colour-coded by environmental class.  We choose to show the version using class as this is the environmental measurement that showed the largest difference in the SAMI galaxies.  To approximate isolated galaxies, we classify any galaxy in EAGLE with $\lmhalo<12$ as isolated.  There is no obvious difference between the environmental classes in EAGLE, although we note that EAGLE has fewer high-mass satellites than SAMI, due to the SAMI cluster sample.  The EAGLE galaxies approximately trace out a sigmoid function.  If we compare the sigmoid fit to the median relation for EAGLE galaxies (black line) to the SAMI relations (grey dotted line) they are in qualitative agreement but somewhat offset (see Table \ref{tab:sigmoid_fit} for fitted values and uncertainties).  The differences are not surprising given the difficulty of measuring absolute stellar population ages and the fact that EAGLE will only be an approximation of the true Universe.

Although not presented in this paper, we note that the correlation analysis results for the full EAGLE data set (not mass-matched to SAMI) give the same qualitative trends as the mass-matched data.

\subsection{simulations at $z>0$}

\begin{figure*}
	\includegraphics[trim=5mm 2mm 5mm 5mm, clip, width=9.2cm]{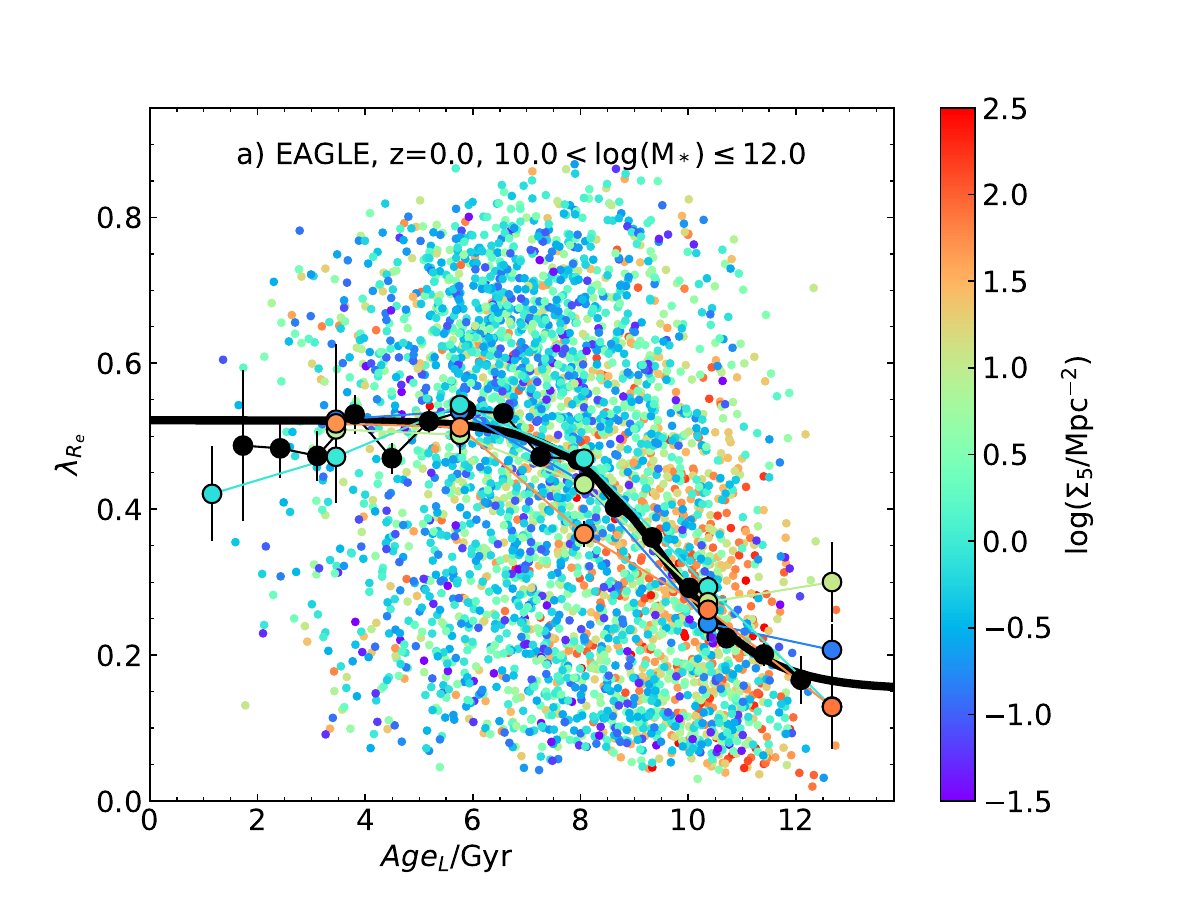}\includegraphics[trim=5mm 2mm 5mm 5mm,clip,width=9.2cm]{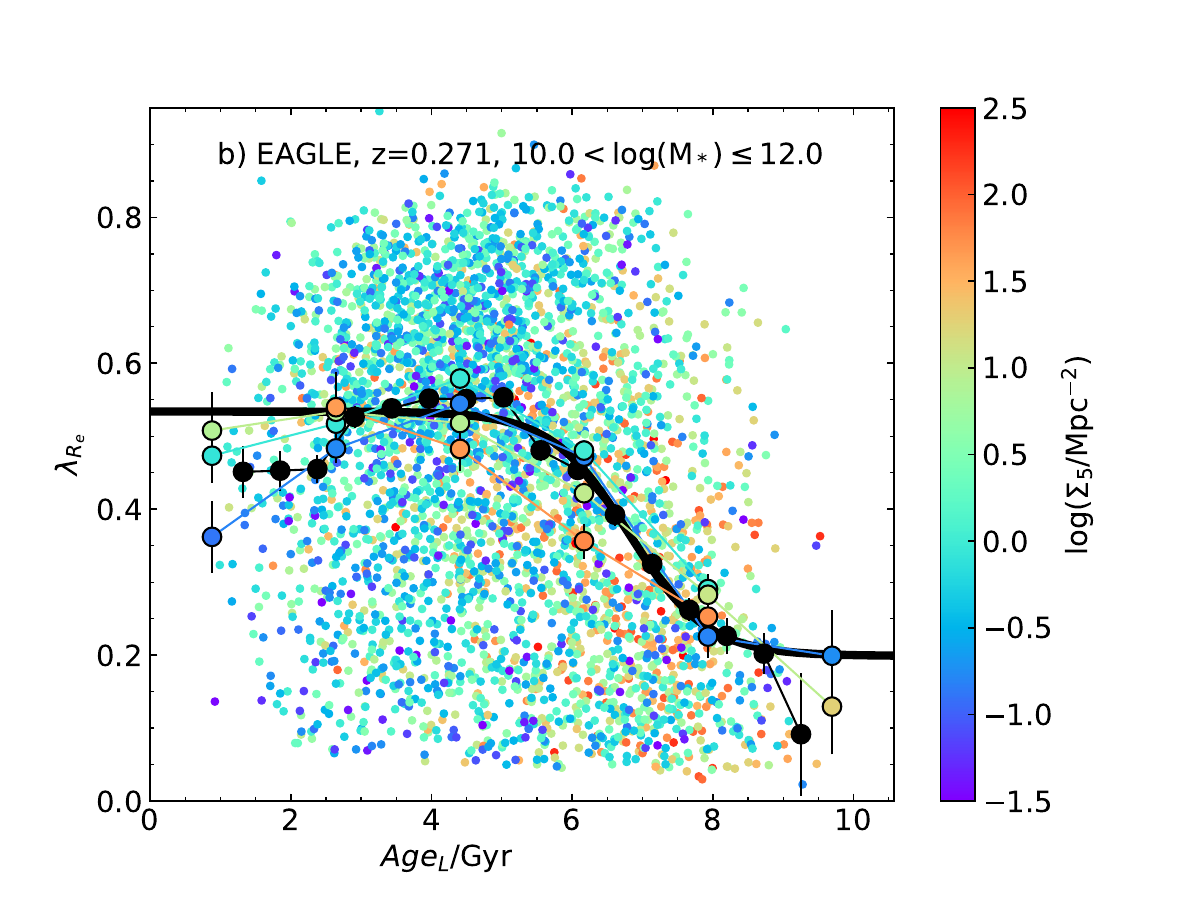}
	\includegraphics[trim=5mm 2mm 5mm 10mm, clip, width=9.2cm]{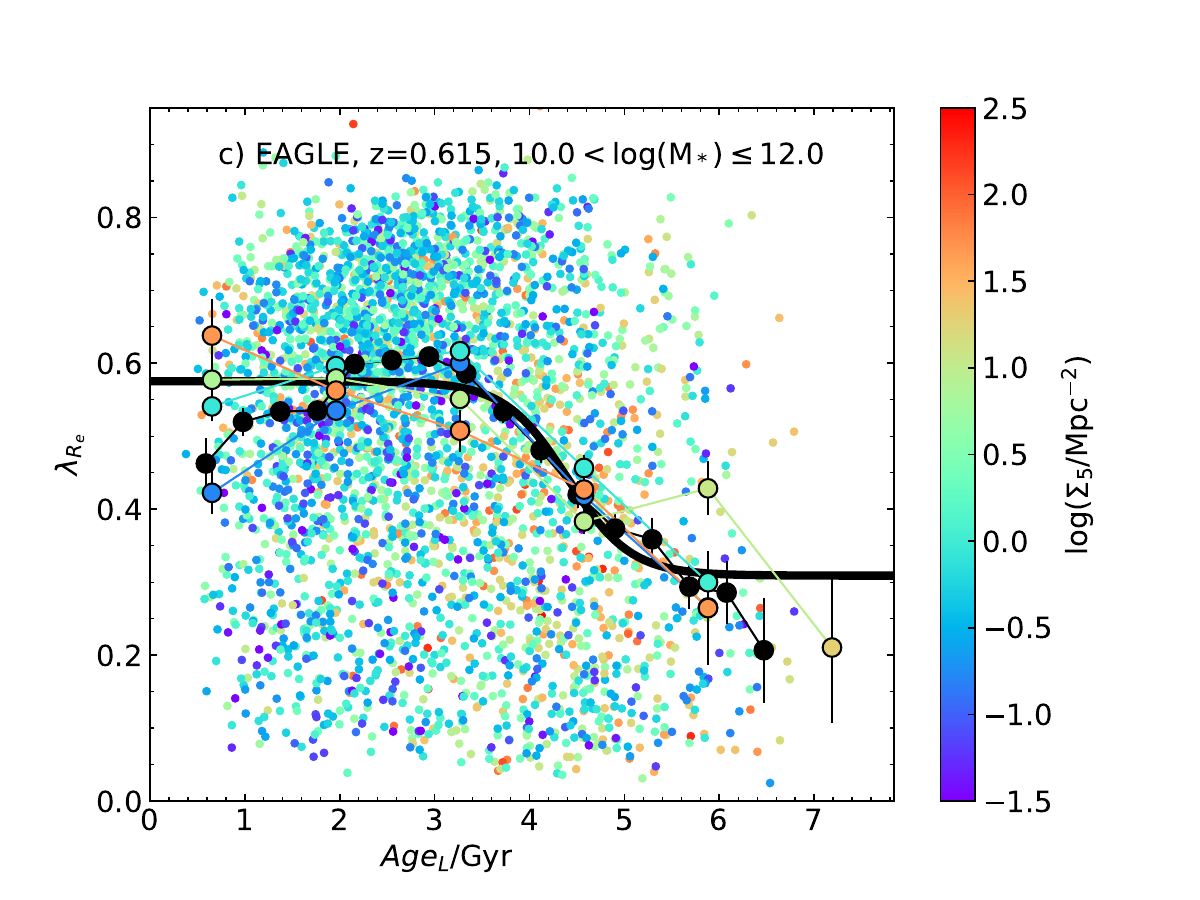}\includegraphics[trim=5mm 2mm 5mm 10mm,clip,width=9.2cm]{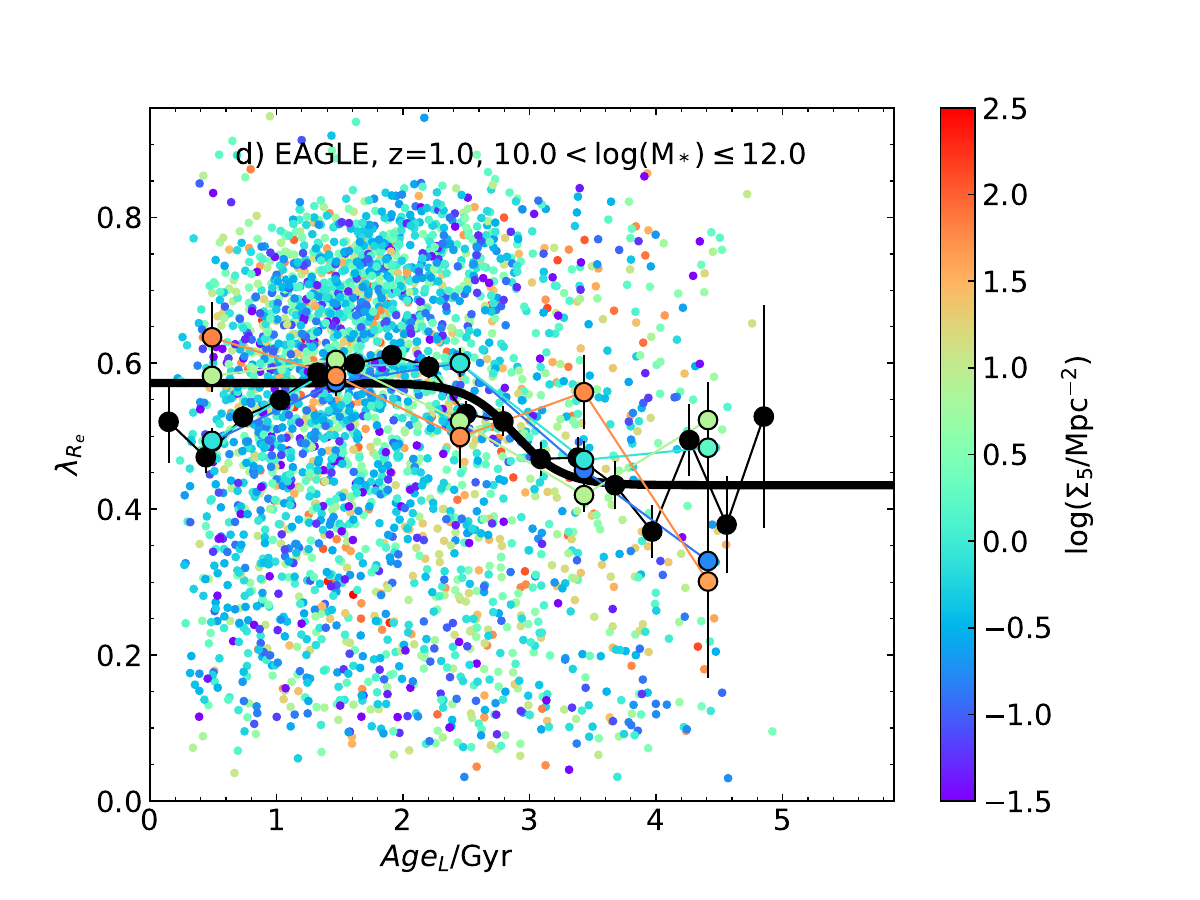}
    \caption{The relation between $\lamre$ and $\agel$ for EAGLE galaxies in 4 different redshift intervals, colour coded by environment using $\Sigma_5$.  In each panel the large back points show the median spin in age bins.  These are fit with a sigmoid function (thick black line).  The large coloured points are the median spin in age bins, but separated into different intervals in environment, using the same colour coding as the smaller points.  Error bars show the uncertainty on the median, calculated using the 68th percentile width of the distribution.  Note that the limits of the x-axis in each panel are different, covering the range from $\agel=0$ to the age of the Universe at the redshift of the snapshot.}
    \label{fig:EAGLE_lam_age_sigma5}
\end{figure*}

Measurements of stellar kinematics are now starting to be made beyond the local Universe \citep[e.g.][]{2021PASA...38...31F,2018ApJ...858...60B}.  Hence, it is valuable to understand how the $\lamre$ vs.\ age relation is expected to change with cosmic time.    We largely focus on the redshift range from $z=0$ to $z=1$ as this is the redshift range that is currently accessible to new observations (although note that \citealt{2023arXiv230806317D} has recently used the James Webb Space Telescope to measure stellar kinematics at higher redshift).

Fig.\ \ref{fig:EAGLE_lam_age_sigma5} shows the $\lamre$ vs.\ $\agel$ trend for four different redshift intervals in EAGLE ($z=0$, 0.271, 0.615, 1.0).  Here we do not mass match to SAMI, as finding sufficient high mass galaxies becomes more challenging at high $z$.  Instead we simply limit the sample to all galaxies with $\lmstar>10$ in each redshift interval.  We analyse the EAGLE data using the same approach that we used for SAMI.  That is, we fit a sigmoid function to the $\lamre$ vs.\ $\agel$ relation.  The $\lamre$ of the youngest EAGLE galaxies does not change very much with $\lambda_0(z=0)=0.522\pm0.001$ and  $\lambda_0(z=1)=0.573\pm0.001$.  While the formal errors on the fitted values of $\lambda_0$ are small, making these changes significant, the overall change is only 0.051.  In contrast the oldest EAGLE galaxies have $\lamre$ that change by a greater amount, with $\lambda_1(z=0)=0.153\pm0.002$ and $\lambda_1(z=1)=0.428\pm0.002$, for a change of 0.276.

The overall pattern that we see is similar to that discussed in various works that study the spin of EAGLE galaxies \citep[e.g.][]{2017MNRAS.464.3850L,2022MNRAS.509.4372L,2021MNRAS.505.2247C}.  That is, young/star-forming galaxies evolve weakly while old/passive galaxies evolve more quickly [e.g.\ see Fig.\ 11 in \citet{2021MNRAS.505.2247C}].  In detail, the rate of evolution depends on factors such as merger history \citep{2022MNRAS.509.4372L}.

In each redshift interval there is relatively little residual environmental impact, but we do see that galaxies at intermediate ages in high density environments (defined by $\lsfnu$) are slightly below the median relation for EAGLE (Fig.\ \ref{fig:EAGLE_lam_age_sigma5}).  This is consistent with the correlation analysis that shows a weak but significant residual correlation between $\lamre$ and $\lsfnu$. 

It is not until $z\sim1$ that there is a substantial change in the $\lamre$-$\agel$ relation.    This predicts that measuring the relation at $z\sim0.3$ using the MAGPI survey \citep{2021PASA...38...31F} should show little evolution compared to $z\sim0$.

\section{Discussion}\label{sec:discussion}

We have analyzed SAMI Galaxy Survey data as a function of stellar mass, age and environment to better understand how these properties impact the dynamical state of galaxies.  The dynamical state is quantified by the spin parameter, $\lamre$.    Using a range of different approaches we consistently find that an age metric is the variable that most strongly correlates with $\lamre$.    The same qualitative result is found whether we use light-weighted age ($\agel$), mass-weighted age ($\agem$) or specific star formation rate [$\lssfrnu$] as our age proxy.

Previous work has shown that galaxy spin is a function of age, mass and/or environment \citep[e.g.][]{2011MNRAS.413..813C,2017ApJ...844...59B,2018NatAs...2..483V,2021MNRAS.508.2307V,2021ApJ...918...84R}.  However, it has not been completely clear how these different relations interact, at least in part due to the smaller sample sizes of previous analyses.  Once we account for $\agel$ or $\lssfrnu$, there is no significant remaining correlation of $\lamre$ with environment parameterized by either $\lsfnu$ or $\lmhalonu$ (across the entire population)  This is not the case if we instead consider $\agem$,  in which case a weak but significant relation between $\lamre$ and environment remains.

If we use class (central, satellite or isolated) as our environmental proxy we see a somewhat different result.  Central galaxies that are older than $\sim7$\,Gyr have significantly (based on K-S test comparisons of the distributions) lower spin than satellites or isolated galaxies.  This result is particularly obvious when we look at the population of galaxies at high stellar mass ($\lmstar>11$).  In this mass range the fraction of slow rotators is known to be significantly higher, and the formation pathway for these galaxies is expected to be different to the bulk of the population, possibly leading to the the environmental trend we see.

Below we will expand on our results in detail, first focusing on the whole galaxy population and the implications of age being the primary driver of galaxy spin. We will then discuss separately the high mass galaxies.

\subsection{Environment drives age, then age drives spin}

The broad empirical view of environmental impact on galaxy formation is clear.  Galaxies in high density environments have lower star formation rates, are redder and are more likely to have early-type morphology.  However, picking apart the causal relationships between these observations is more complex.  The ATLAS3D survey of local early-type galaxies was the first to show that environment appeared to play a role in the dynamical structure of galaxies \cite{2011MNRAS.416.1680C} and this result was supported by a number of other measurements \citep[e.g.][]{2013MNRAS.429.1258D,2014MNRAS.443..485F}.  The following generation of IFS surveys \citep[e.g.][]{2017ApJ...844...59B,2017MNRAS.471.1428V} showed that the stellar mass was important, but there remained evidence for a separate environmental trend \citep{2021MNRAS.508.2307V}.

The natural implication of our result that age correlates most strongly with $\lamre$ is that environment-$\lamre$ is a secondary correlation, caused by the fact that age is influenced by environment.  More explicitly, quenching processes due to environment (e.g.\ ram pressure stripping) lead to the suppression of star formation and as a result galaxies in dense environments are older, as they contain fewer recently formed stars.  This is true across a range of environments, both in clusters and groups, although the quenching efficiency (or timescale) seems to depend on the density or halo mass of the environment \citep[e.g.][]{2017ApJ...847..134K,2019ApJ...873...52O,2022MNRAS.516.3411W,2022ApJ...941....5J}.  Our picture is also supported by the fact that we find that $\lssfrnu$ and $\agel$ are better correlated with spin than $\agem$.  The value of $\lssfrnu$ is obviously a good measure of the level of current star formation, while $\agel$ is a reasonable proxy for the time since the last substantial episode of star formation.  One caveat to note is that our $\lamre$ measurement is also light weighted.  If we could make a mass-weighted $\lamre$ measurement it is possible we would find a different relation between $\lamre$ and age.  However, above we found that at least in the EAGLE simulations a mass-weighted estimate of $\lamre$ was still most strongly correlated with light-weighted age or sSFR.

A particularly interesting aspect of $\agel$ or $\lssfrnu$ being the primary driver of $\lamre$ is that these are quantities that depend on how the gas content of galaxies evolves over cosmic time.  They have no direct relationship to processes that modify the dynamics of stars already formed.  Put another way, the current value of $\lamre$ appears to be more related to gas processes (i.e. processes that remove gas or quench star formation) than purely gravitational processes.  There are a number of gravitational processes in high density environments that have been proposed to modify the morphology of galaxies \citep[e.g.][]{1996Natur.379..613M,2011MNRAS.415.1783B}.  It appears that these gravitational processes cannot be dominant in setting $\lamre$, as galaxies of the same age have, on average, the same $\lamre$ irrespective of the environment they inhabit.  That said, we note an apparent split by environmental class for massive galaxies that we will discuss in Section \ref{sec:dis_highmass}.

Above we make the case that quenching and/or aging leads to a change in galaxy dynamics.  However, one could argue that the strong correlation we see between age and spin actually goes in the other direction.  That is, a higher dispersion central component in a galaxy could lead to lower star formation rate via morphological quenching \citep{2009ApJ...707..250M}.  There is evidence that local early-type galaxies have lower star formation efficiency, at least in part due to their steep inner rotation curve, which provides a sheer that helps stabilize giant molecular clouds (GMCs) against collapse \citep{2014MNRAS.444.3427D,2018MNRAS.475.1791C,2022MNRAS.512.1522D}.  However, it is also recognised that morphological quenching cannot generate all of the passive population \citep[e.g.\ see discussion by ][]{2013MNRAS.432.1914M}.  Indeed, \citet{2021MNRAS.503.4992F} find that there is little change in galaxy spin until galaxies are at least 1 dex below the star-forming main sequence, suggesting that the reduction in star formation cannot be driven by structural change.  Similarly, \citet{2019MNRAS.485.2656C} show that during their quenching phase satellites have little change in their kinematics.   Thus, while morphological quenching may contribute to our observed relation between age and spin, it is unlikely to dominate.  That said, future investigations examining the spin-age relation as a function of radius (and therefore velocity gradient) would be valuable for clarifying this point.

Another consideration regarding the connection between structure and age is evidence that stellar population age correlates better with stellar mass surface density (i.e. $\sim M_*/\re^2$) than just mass \citep[e.g.][]{2017MNRAS.472.2833S,2018ApJ...856...64B,2020ApJ...898...62B}.  We explore this by replacing $M_*$ with $M_*/\re^2$ (and also  $M_*/\re$, i.e. approximately gravitational potential) and re-running our partial correlation analysis (from Section \ref{sec:correlation}).   When we do this, the qualitative results are unchanged. $\lamre$ still correlates best with $\agel$ ($r=-0.451$ and $-0.462$ when using $M_*/\re$ and $M_*/\re^2$ respectively), while the partial correlations between $\lamre$ and $\log(M_*/\re)$ or $\log(M_*/\re^2)$ are not significant ($r=-0.001$ and $r=-0.010$ respectively).  Thus, while compact galaxies are older, this does not appear to drive low spin (at least for the bulk of the population).

Although our observations suggest that the relationship between spin and environment is indirect, it is worth noting two caveats.  First, the environment we observe is a snapshot at the current epoch, while a galaxy's properties will be influenced by the integrated environmental effects across it's entire history.  At the very least, this will add intrinsic scatter to relations between environment and other galaxy properties and mask direct relationships.  It could be that a property such as $\agel$ is a better proxy for the integrated environmental impact on a galaxy than the instantaneous environment.

Our second caveat is that from tidal torque theory \citep{1951pca..conf..195H,1969ApJ...155..393P} we expect the spin of a galaxy to be related to at least its early environment.  Simulations show that dark matter haloes are aligned with large-scale structure, such as the filaments and walls of the cosmic web \citep[e.g.][]{2013ApJ...762...72T}.  Further, simulated galaxies are also predicted to align with large-scale structure \citep{2018MNRAS.481.4753C}.  However, late time evolution is expected to wash out some of this alignment effect.  Observations are now starting to find evidence that the direction of galaxy spin is indeed aligned with large-scale structure \citep[e.g.][]{2020MNRAS.491.2864W,2022MNRAS.516.3569B,2023MNRAS.526.1613B}.  While the direction of spin does seem to align with large-scale structure, whether the amplitude of spin (e.g. $\lamre$) does so is less clear.    \citet{2018MNRAS.481..414G} compares the spin parameter of simulated haloes across nodes, filaments and walls.  They use a spin parameter defined as $\lambda = J/(\sqrt{2}MVR)$, where $J$ is the amplitude of angular momentum, $M$ is the halo mass and $V$ is the halo circular velocity at radius $R$ \citep[following][]{2001ApJ...555..240B}, which is similar to the dimensionless spin parameter defined by \citet{1969ApJ...155..393P}.  \citet{2018MNRAS.481..414G} find that haloes in filaments and walls have a typical value of $\lambda=0.035$, but that this is reduced to $\lambda=0.02$ for nodes in large-scale structure.  It is uncertain whether this halo spin amplitude relation with large-scale structure also translates to galaxies.  It is plausible that baryonic physics, such as the cooling of gas and/or feedback are the dominant processes in setting spin on galaxy scales, and that is why we do not see any residual relation with environment.  The question of spin across the cosmic web will be discussed in detail by Barsanti et al (in prep).

\subsection{Why do old galaxies have low spin?}

If age is the primary driver of galaxy spin, with little dependence on environment or mass (we will discuss massive slow rotators separately below), this raises the question of {\it why} age drives spin.  Several possible  explanations are:
\begin{itemize}
\item {\it Internal or secular dynamical heating:} once a galaxy gets rid of its gas and stops forming new stars in a thin disc, internal dynamical heating will have a greater impact to reduce spin.
\item {\it Progenitor bias:} stellar populations in the early Universe are born hotter, and if quenching happens earlier there is less time to build a thin disc with high spin.
\item {\it Dry mergers:} the progenitors of older galaxies quenched earlier, making it more likely that they experience dry mergers. 
\item {\it Disc fading:} as the stellar populations fade over time, a younger disc becomes less dominant, increasing the importance of any older dispersion-dominated bulge.
\end{itemize}
The last of these items has been discussed in detail by \cite{2021MNRAS.505.2247C}, who built simulations to estimate the impact of disc fading on stellar kinematics.  They concluded that while $\lamre$ is reduced by disc fading, it is not sufficient to explain the difference in $\lamre$ seen between star-forming and passive disc galaxies in the local Universe.  As a result, we will largely focus our discussion on the first three points above, noting the strong possibility that multiple processes may contribute.

At least in the local Universe the vast majority of stars are formed in dynamically cold discs.  For example, the molecular gas that stars are born from typically has velocity dispersions of $\sim10$\,\kms\ or less \citep[e.g.][]{2016AJ....151...15M,2018ApJ...860...92L}.  While star formation is ongoing the young cold disc will tend to keep the value of 
$\lamre$ relatively high.  Once star formation ceases, dynamical heating will tend to increase the velocity dispersion of stars.  In the Milky Way there is clear observational evidence for an age-velocity dispersion relation \citep[e.g.][]{2021MNRAS.506.1761S}.  Comparison of this to idealized simulations \citep[e.g.][]{2016MNRAS.459.3326A} suggests that much (although not all) of the Milky Way's  age-velocity dispersion relation could be due to stars scattering off of GMCs or the influence of bars and spiral arms.  If GMCs dominate scattering, then the scattering should reduce substantially once a galaxy quenches and has a lower molecular gas content.  As a result, it is not obvious that such scattering could drive a spin-age relation in quenched galaxies with older stellar populations.  Instead, bars may need to be the primary mechanism (detailed simulations of the impact of this would be valuable).  High resolution cosmological simulations \citep[e.g.][]{2021MNRAS.503.5826A,2023arXiv230803566Y} find that while late-time secular effects do also lead to an age-velocity dispersion relation, early mergers and other accretion events also provide heating.  Thus, while secular heating may play a role, it is not obvious whether it is the dominant mechanism to increase $\lamre$ in already quenched galaxies.

The second of our options above concerns progenitor bias.  Observationally there is now strong evidence that the ionized gas velocity dispersion ($\sigmagas$) in high redshift galaxies is substantially higher than in typical local galaxies \citep[e.g.][]{2015ApJ...799..209W,2018MNRAS.474.5076J,2019ApJ...880...48U}.  While in the local Universe $\sigmagas$ measured from H$\alpha$ is typically $\sim15-20$\kms \citep[e.g.][]{2020MNRAS.495.2265V,2020MNRAS.498.5885B}, this more than doubles by $z\sim2$ to $\sim45$\kms\ \citep{2019ApJ...880...48U}.  Assuming that the ionized gas in high redshift galaxies has similar kinematics to the newly formed stars, the dispersion evolution implies that discs at high redshift are more turbulent.  Thus older stellar populations are expected to be dynamically hotter.  Local examinations of individual S0 galaxies seem to agree with this picture.  For example, \citet{2019MNRAS.487.3776P} find that the oldest components of NGC3115 have dispersions that are 2-3 times higher than the youngest components, in qualitative agreement with the redshift evolution of dispersion.

The transition from high to low $\lamre$ is centred around $x_0=10.0\pm0.8$\,Gyr when fitting a sigmoid function to the relationship between $\lamre$ and light-weighted age (Section \ref{sec:full_sample}).  However, as seen in Fig.\ \ref{fig:lam_age_sigma5}, the transition is gradual, starting at $\agel\sim5$\,Gyr.   If we assume that the light-weighted age is a reasonable proxy for the look-back time when the last major star formation episode occurred, then the ages correspond to a redshift of $z\sim0.6$ (5\,Gyr) and $z\sim1.7$ (9.9\,Gyr).

\begin{figure*}
	\includegraphics[trim=5mm 2mm 5mm 5mm, clip, width=14cm]{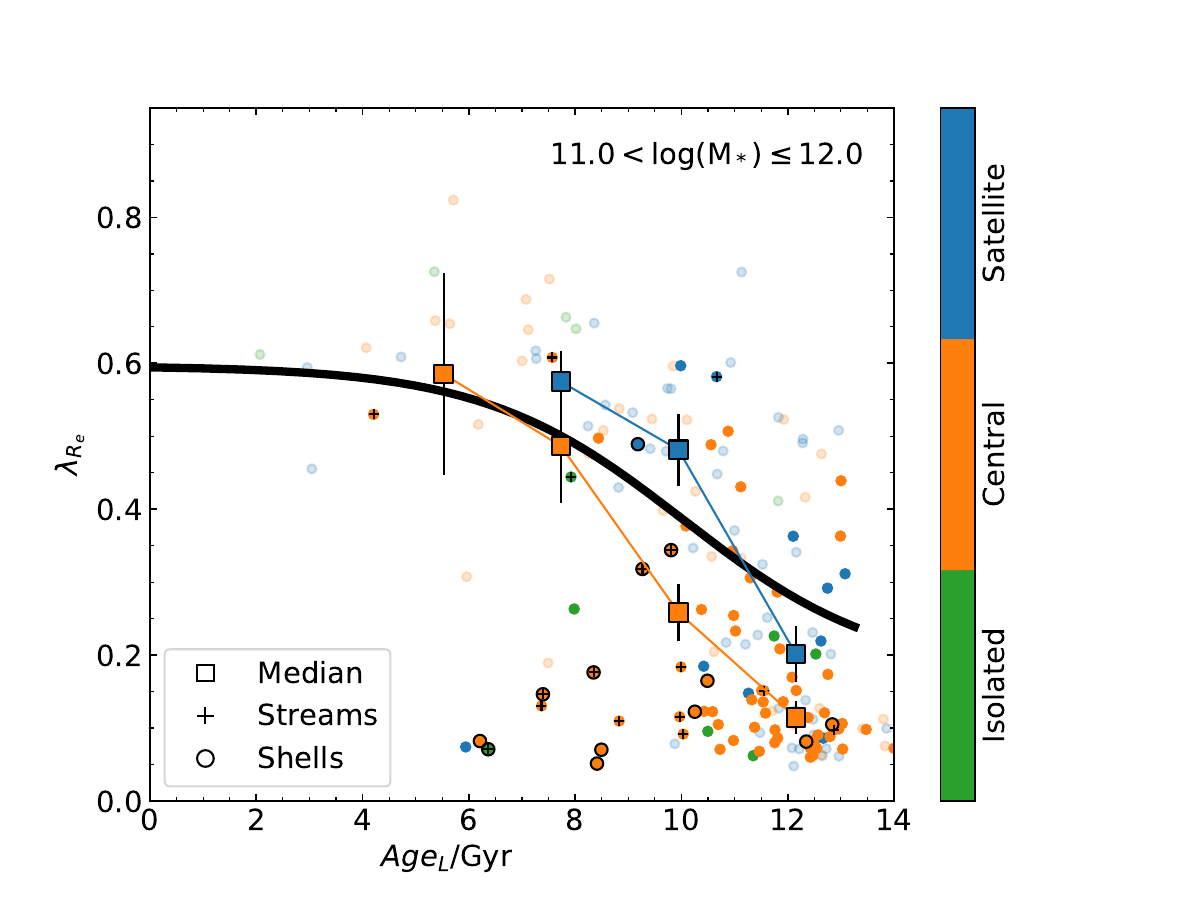}
    \caption{The relation between $\lamre$ and $\agel$ for SAMI galaxies at $\lmstar>11$, indicating galaxies that show evidence of tidal streams (black crosses) or shells (black circles) in deep imaging \citep{2024arXiv240202728R}.  Points are colour-coded by environmental class with the galaxies analysed by Rutherford et al.\ shown by bright points and those that were not analysed being faint.  The large square points show the median spin in age bins and the black solid line is the median relation for all galaxies in our full sample (identical to that shown in Fig. \ref{fig:lam_age_sigma5}g).}
    \label{fig:shells}
\end{figure*}

\citet{2019ApJ...880...48U} find a relation $\sigmagas = (23.3\pm4.9) + (9.8\pm3.5) z$ for the mean ionized gas velocity dispersion as a function of redshift.  The zero-point at $z=0$ is likely a little high compared to some recent work \citep[e.g.][]{2020MNRAS.495.2265V}, but this model broadly captures the evolution of gas dispersion.  The model suggests that gas dispersions are $\sim29$\,\kms\ at $z\sim0.6$ and $\sim40$\,\kms\ at $z\sim1.7$.  The increased dispersion at birth of these stars will naturally contribute to the lower $\lamre$ of older galaxies.

Simulations by \citet{2022MNRAS.512.3806V} show that isolated galaxies with high gas fractions have stars born with higher dispersion.  The high gas fractions at high redshift then naturally point to higher stellar dispersion.  \citet{2023MNRAS.524.4346J} find that evolution in gas velocity dispersion is similar in EAGLE to observations \citep[e.g.][]{2019ApJ...880...48U}.  High resolution cosmological simulations by the FIRE-2 team \citep{2023ApJS..265...44W} show that in Milky Way like galaxies, stars formed at high redshift are born dynamically hot \citep{2022arXiv221003845Y}.  \citet{2023MNRAS.525.2241H} uses suites of hydrodynamic simulations to show that increased mass concentration seems to be the main driving factor in forming a thin disc.  In the context of our results, galaxies with old stellar populations will have built most or all of their stars  before the inner mass profile was sufficiently concentrated, which results in more turbulent cold gas flows and high stellar velocity dispersion at birth.

The third route that we list above to reduce spin is via merging.  The remaining stellar mass in a rotating disc after a merger depends on a number of parameters, including mass ratio, orbital configuration and gas fraction \citep[e.g.\, see the detailed discussion by][]{2009ApJ...691.1168H}.  In particular, even with 1-to-1 mass ratios, gas rich mergers do not substantially reduce the fraction of stars in a rotating disc, as the gas allows for the reformation of a disc.  This result is borne out in cosmological simulations by \citet{2017MNRAS.464.3850L}, who examine the angular momentum in galaxies within the EAGLE simulations \citep{2015MNRAS.446..521S}, including both the specific angular momentum in stars, $j_{\rm stars}$ and a spin parameter of the form $\lambda'_{\rm stars}=j_{\rm stars}/M^{2/3}_{\rm stars}$.  They find that lower angular momentum (or spin) is driven by either early quenching or dry mergers.  Galaxies with continuing star formation tend to maintain or increase their spin, even in the presence of major mergers.  On the other hand, galaxies that quench early have consistently lower spin, and the difference is made greater by dry mergers [see Fig. 13 of \cite{2017MNRAS.464.3850L}].  

While the older galaxies in the SAMI sample have lower spin than young galaxies, most still fall within a parameter space that is consistent with regular oblate rotators and would typically be classified as lenticular or S0 galaxies.  This population is therefore unlikely to have been formed by dry major mergers, as the mergers would have destroyed their discs [although we note that some S0s may be formed during gaseous mergers \citep{2020MNRAS.498.2372D,2021MNRAS.508..895D}].   Dry minor merging would typically allow the disc to survive, while decreasing spin.  However, the ability to merge is a function of environment, with mergers becoming more difficult in cluster environments.  Given that there is no significant difference between the spin-age relations as a function of halo mass [at least at $\lmstar<11$, e.g. Fig.\ \ref{fig:lam_agel_mass_halo}], it is not clear whether dry minor merging can be a significant contribution to reducing spin within 1$R_e$.

The work of \citet{2017ApJ...837...68C}, looking at simulated galaxy clusters also shows that while mergers can spin down galaxies, even galaxies without major or minor mergers will spin down over cosmic time.  The non-merger route seems particularly important at lower masses ($\lmstar<10.5$ in their simulations).  \citet{2017ApJ...837...68C} suggest that the cause of this spin down could be fly-by interactions or other tidal effects.  However, the degree of such interactions should be dependent on environment and we find that SAMI galaxies in cluster mass halos (e.g.\ red points in Figs.\ \ref{fig:lam_age_sigma5}d, e, and f) are not at significantly lower spin than the rest of the population, once age is accounted for.  At this point it is worth noting that our measurements are for spin flux-weighted within 1 $R_e$.  At larger radii, where the stars are less bound, gravitational tidal effects likely become more important.

In summary, it seems most likely that the age-spin relation we find is driven by progenitor bias, with stars being born dynamically hotter in the early Universe.  While other physical processes may contribute, there are various reasons to rule them out as dominant.  \citet{2021MNRAS.505..991C} showed that the difference in spin between current passive and star-forming discs is too large to be accounted for by disc fading.  Internal secular heating is possible, but if this is driven by GMCs, it could be less important in quenched galaxies. Instead bar-related processes may be needed. For the bulk of the population, dry major mergers cannot be the route, as most passive galaxies still have a significant disc.  Dry minor mergers may contribute, but we do not find the environmental trend that might be expected with these.

\subsection{What is special about the spin of high mass galaxies?}\label{sec:dis_highmass}

Up to this point, our discussion has focussed on the bulk of the galaxy population, that is dominated by oblate rotators.    However, at high mass ($\log(M_*/M_{\odot})\gtrsim11$) there is an increasing population of galaxies that cannot be modeled as oblate rotators that we call slow rotators \citep[e.g.][]{2007MNRAS.379..401E,2011MNRAS.413..813C}.  This population is small enough that when our analysis is carried out across the entire SAMI sample they do not strongly impact the correlation analysis.  However, we start to see different behaviour when we split the sample by mass and examine the galaxies at $\lmstar>11$.  

For the high mass population we still find a strong relationship between spin and age (or sSFR).  There is no significant environmental trend when using $\lsfnu$, which is consistent with the picture found by \citet{2024arXiv240203676V}, who use logistic regression to find the most important parameters that determine whether a galaxy is a slow rotator or not.  They find that mass and SFR, along with size and ellipticity, are the most important parameters and once SFR is included (similar to our use of age) there is no residual environmental trend (using $\lsfnu$).  When we use other environmental metrics the result is somewhat different.
When we examine the spin of high mass galaxies as a function of environmental class, centrals consistently have lower spin than satellites of the same mass for all age metrics we use ($\agel$, $\agem$, $sSFR$).  

The environmental impact on high-mass galaxies can be most clearly seen in Figs.\  \ref{fig:lam_agel_mass_sigma5}d, \ref{fig:lam_agel_mass_halo}d and \ref{fig:lam_agel_mass_satcent}d.  There are two specific trends that it is important to recognize.  The first is that the high-mass galaxies with low spin and the oldest ages usually sit in the very highest density environments as defined by $\lsfnu$ (orange points in \ref{fig:lam_agel_mass_sigma5}d).  The second is that high-mass galaxies with low spin, but intermediate ages ($\agel\sim6-11$\,Gyr) are preferentially central galaxies (orange points in \ref{fig:lam_agel_mass_satcent}d) in intermediate mass halos (cyan points in \ref{fig:lam_agel_mass_halo}d). 

There is now good evidence from both observations \citep[e.g.][]{2014MNRAS.444.3986R} and simulations \citep[e.g.][]{2010ApJ...725.2312O,2020MNRAS.497...81D} that galaxy merging becomes more important for stellar mass growth as galaxy mass increases.   In particular, the fraction of ex-situ star formation becomes greater than $\sim0.5$ above $\lmstar\sim11$ \citep[albeit, with some scatter depending on the particular simulations; see Fig. 2 of][] {2020MNRAS.497...81D}.  The importance for stellar mass growth of mergers at high mass is consistent with the expected formation pathways of slow rotators from simulations, where most undergo merging \citep{2022MNRAS.509.4372L}.

The difference in the $\lamre-\agel$ relation between high-mass centrals and satellites (e.g.\ Fig.\ \ref{fig:lam_agel_mass_satcent}d) is qualitatively consistent with being driven by mergers.  Central galaxies sit in a preferred location within halos, which allows satellites to more efficiently merge with them via dynamical friction.  On the other hand, satellites are less likely to merge with other satellites \citep[e.g.][]{2020MNRAS.497...81D}, particularly in cluster mass halos where relative velocities are high.  In Fig.\ \ref{fig:lam_agel_mass_halo}d there is a population of high-mass galaxies with low spin ($\lamre\lesssim0.2$) with $\agel\simeq7-11$\,Gyr that are mostly centrals in group mass halos ($\lmhalo\simeq13-14$).  While some centrals in similar mass halos do lie close to the median $\lamre-\agel$ relation, there is a clear excess at $\lamre\lesssim0.2$.  Given that merging is more efficient in group-like environments (particularly for central galaxies), this supports a merger route for the formation of these slow rotators.  \citet{2023MNRAS.521.2671S} use dynamical modelling of a sub-sample of SAMI early-type galaxies to show that central galaxies have higher radial anisotropy, broadly consistent with a merger hypothesis.

\citet{2024arXiv240202728R} examine early-type galaxies in the SAMI sample to identify low-surface-brightness streams and shells using Hyper Suprime Cam imaging \citep{2018PASJ...70S...4A}.  They find that there is a preference for galaxies with shell-like features to have lower spin when separated by stellar population age.  Fig.\ \ref{fig:shells} shows the age-spin relation at high mass, highlighting the galaxies that are found by Rutherford et al.\ to have tidal streams (crosses) or shells (circles).  While not all galaxies in our sample are analysed by Rutherford et al.\ (they focus on early-type galaxies in the GAMA regions), all of the low spin centrals at ages $\sim6-10$\,Gyr that are examined show evidence of tidal features.  The features are much less common for galaxies with older ages $\gtrsim10$\,Gyr.   If mergers formed these older galaxies, the mergers may have happened early enough that the tidal features are no longer visible.

At intermediate age ($\agel\sim8-11$\,Gyr) almost all high-mass satellites have high spin (at or above the median line in the $\lamre-\agel$ relation).  The majority of these high-mass satellites come from the SAMI cluster sample.  The low probability of these galaxies merging with another satellite is likely to be the reason for their high spin.

At the oldest ages ($\gtrsim12$\,Gyr) there is an excess of massive low $\lamre$ galaxies for both centrals and satellites.  This result is natural for central galaxies, due to the expected higher rates of merging for these galaxies.  For the satellites, it is possible that mergers occurred when the galaxies were the centrals of groups, prior to infall into the cluster.  However, it is worth noting that in EAGLE simulations approximately half of slow rotators that are satellites at $z=0$  experience mergers when they are already satellites, generally while in the outskirts of haloes, where the galaxy-to-galaxy velocity dispersion is smaller \citep{2022MNRAS.509.4372L}.  Given that slow rotator formation generally only happens in dry mergers \citep{2018MNRAS.473.4956L}, the extremely old light weighted age in these objects means that the relevant mergers could have taken place very early, when merger rates were much higher.

More broadly, the requirement that quenching needs to happen before merging to make slow rotators \citep{2018MNRAS.473.4956L} means central galaxies need to be {\it mass quenched}.  The quenched fraction of central galaxies is a strong function of mass \citep[e.g.][]{2014MNRAS.441..599B}, with the central galaxy quenched fraction becoming greater than 0.5 above $\log(M_*/M_\odot)\sim10.5$ in the local Universe.  At higher redshift the quenched fraction reduces \citep[e.g.][]{2017MNRAS.472.2054P}, meaning that the dry merger rate will not increase in the same way as the total merger rate \citep[e.g.\ see Fig.\ 2 of][]{2018MNRAS.473.4956L}.  Given that the likelihood of undergoing a dry merger will increase with the length of time galaxies are passive, it is unsurprising that for old high-mass galaxies we see a high fraction of low spin galaxies.

\section{Conclusions}\label{sec:conc}

We have analysed the full SAMI Galaxy Survey sample \citep{2021MNRAS.505..991C} to better understand the drivers of galaxy spin, $\lamre$.  Our conclusions are as follows:

\begin{enumerate}
    \item Using a linear partial correlation analysis, age proxies (either light-weighted age, mass-weighted age or specific star formation rate) consistently show the strongest correlation with spin, $\lamre$.  Correlations with environment or mass are always much weaker than with age.
    \item Analysing the full SAMI sample when correcting for light-weighted age ($\agel$) or specific star formation rate ($sSFR$) there is no residual correlation between $\lamre$ and environment, or $\lamre$ and stellar mass.  Mass-weighted age ($\agem$) is less well correlated with $\lamre$ and in this case there are residual correlations with mass and environment.
    \item The median trends in the  $\lamre$-age plane are well modelled by a sigmoid function, with an RMS scatter around the relation of 0.15 in $\lamre$.  This scatter is relatively small given that we are using projected $\lamre$ values without correcting for the impact of inclination.
    \item The lack of environmental correlation with spin (other than through a secondary correlation with age or specific star formation rate) is suggestive that environmentally dependent gravitational processes (such as galaxy-galaxy interactions) are not important, at least for stellar kinematics measured within $1\re$.
    \item One residual environmental trend that does appear in our data is that central galaxies have lower spin than satellites at high stellar mass, $\lmstar\gtrsim11$.  This is particularly obvious at intermediate ages, $\agel\simeq7-11$\,Gyr and could plausibly be driven by merging being more commonplace for centrals, and provides a route to build the slow rotator population.
    \item Stellar kinematics measured from EAGLE simulations show a similar qualitative spin-age relation to our observational data.  These simulations predict that the spin of young galaxies will not evolve strongly up to at least $z=1$.  However, the population of the oldest galaxies has a significant evolution from median $\lamre\sim0.45$ at $z=1$ to $\lamre\sim0.15$ at $z=0$.
    \item Within the EAGLE simulations both light-weighted and mass-weighted estimates of $\lamre$ show stronger correlations with $\agel$ or sSFR than $\agem$.   This suggests that our use of light-weighted $\lamre$ is not the cause of the age-spin correlation we find. 
\end{enumerate}

Given that age seems to be the strongest predictor of $\lamre$, a promising approach to further understanding the drivers of galaxy spin will be to measure the spin-age relation for galaxies at higher redshift to examine the evolution of the relation.  High quality stellar kinematics of high-redshift galaxies is challenging, but is now starting to be possible with projects such as the LEGA-C survey \citep{2018ApJ...858...60B} using slits at $z\sim1$, or the Middle-Ages Galaxy Properties with Integral Field Spectroscopy \citep[MAGPI;][]{2021PASA...38...31F} survey currently proceeding on the Very Large Telescope with MUSE at $z\sim0.3$.  While gas kinematics at high redshift is now routine, stellar kinematics is not.  The advent of ELTs should revolutionize this, with instruments such as HARMONI on ELT \citep{2010SPIE.7735E..2IT} and GMTIFS \citep{2012SPIE.8446E..1IM} and MANIFEST \citep{2010SPIE.7735E..68S} on the Giant Magellan Telescope.

\section*{Acknowledgements}

The SAMI Galaxy Survey is based on observations made at the Anglo-Australian Telescope. The Sydney-AAO Multi-object Integral field spectrograph (SAMI) was developed jointly by the University of Sydney and the Australian Astronomical Observatory. The SAMI input catalogue is based on data taken from the Sloan Digital Sky Survey, the GAMA Survey and the VST ATLAS Survey. The SAMI Galaxy Survey is supported by the Australian Research Council Centre of Excellence for All Sky Astrophysics in 3 Dimensions (ASTRO 3D), through project number CE170100013, the Australian Research Council Centre of Excellence for All-sky Astrophysics (CAASTRO), through project number CE110001020, and other participating institutions. The SAMI Galaxy Survey website is \href{http://sami-survey.org/}{http://sami-survey.org/}.

GAMA is a joint European-Australasian project based around a spectroscopic campaign using the Anglo-Australian Telescope. The GAMA input catalogue is based on data taken from the Sloan Digital Sky Survey and the UKIRT Infrared Deep Sky Survey. Complementary imaging of the GAMA regions is being obtained by a number of independent survey programmes including GALEX MIS, VST KiDS, VISTA VIKING, WISE, Herschel-ATLAS, GMRT and ASKAP providing UV to radio coverage. GAMA is funded by the STFC (UK), the ARC (Australia), the AAO, and the participating institutions. The GAMA website is \href{http://www.gama-survey.org/}{http://www.gama-survey.org/}. 

JvdS acknowledges support of an Australian Research Council Discovery Early Career Research Award (project number DE200100461) funded by the Australian Government.   JJB acknowledges support of an Australian Research Council Future Fellowship (FT180100231). SKY acknowledges support from the Korean National Research Foundation (2020R1A2C3003769, 2022R1A6A1A03053472).  AR acknowledges the receipt of a Scholarship for International Research Fees (SIRF) and an International Living Allowance Scholarship (Ad Hoc Postgraduate Scholarship) at The University of Western Australia.
SMS acknowledges funding from the Australian Research Council (DE220100003).
FDE acknowledges support by the Science and Technology Facilities Council (STFC), by the ERC through Advanced Grant 695671 ``QUENCH'', and by the UKRI Frontier Research grant RISEandFALL.

\section*{Data Availability}

The SAMI data used in this paper, including kinematic measurements, are included in SAMI Data Release 3 \citep{2021MNRAS.505..991C} available via Australian Astronomical Optics' Data Central, https://datacentral.org.au/.



\bibliographystyle{mnras}
\bibliography{bibliographies} 




\appendix

\section{Correlation analysis for other sub-samples}\label{app:a}

In this appendix we include the tabulated results of correlation analysis not presented in the main paper.  This mostly consists of sub-samples used to test the robustness of the result.   We list correlation analysis results for the restricted sample ($\lmstar>10$ and no slow rotators) in Table \ref{tab:corr_four_tests}.  The results from the restricted sample split into GAMA regions only and clusters only are listed in Tables \ref{tab:corr_four_restricted_gama} and \ref{tab:corr_four_restricted_cluster} respectively.

\begin{table}
	\centering
	\caption{The results of our correlation analysis between 4 parameters: $\lmstarnu$, $\lsfnu$, $\lamre$ and an age proxy.  The format is the same as Table \ref{tab:corr_four_full}, but lists results using the restricted sample ($\lmstar>10.0$ and no slow rotators).  Correlations with $P>0.01$ are in bold.}
	\label{tab:corr_four_tests}
	\begin{tabular}{ccrrrr} 
\hline
& & \multicolumn{2}{c}{full correlation} &\multicolumn{2}{c}{partial correlation}\\
A & B & $r$ & $p$-value & $r$ & $p$-value \\
\hline
$\lmstarnu$ & $\agel$  &  0.431 & 1.20e-52 &   0.410 & 2.33e-47 \\
$\lmstarnu$ & $\lsfnu$ &  0.071 & {\bf 1.65e-02} &  -0.093 & 1.69e-03 \\
$\lmstarnu$ & $\lamre$ & -0.179 & 1.20e-09 &   0.073 & {\bf 1.42e-02} \\
$\agel$  & $\lsfnu$ &  0.343 & 9.47e-33 &   0.320 & 1.60e-28 \\
$\agel$  & $\lamre$ & -0.538 & 2.63e-86 &  -0.504 & 2.22e-74 \\
$\lsfnu$ & $\lamre$ & -0.156 & 1.12e-07 &   0.042 & {\bf 1.60e-01} \\
\hline
$\lmstarnu$ & $\agem$  &  0.445 & 1.84e-56 &   0.414 & 2.75e-48 \\
$\lmstarnu$ & $\lsfnu$ &  0.071 & {\bf 1.65e-02} &  -0.040 & {\bf 1.83e-01} \\
$\lmstarnu$ & $\lamre$ & -0.179 & 1.20e-09 &  -0.020 & {\bf 5.08e-01} \\
$\agem$  & $\lsfnu$ &  0.234 & 1.13e-15 &   0.191 & 8.02e-11 \\
$\agem$  & $\lamre$ & -0.371 & 1.73e-38 &  -0.309 & 1.51e-26 \\
$\lsfnu$ & $\lamre$ & -0.156 & 1.12e-07 &  -0.078 & 8.79e-03 \\
\hline
$\lmstarnu$ & $\log(sSFR)$ & -0.303 & 5.41e-26 &  -0.251 & 5.40e-18 \\
$\lmstarnu$ & $\lsfnu$ &  0.072 & {\bf 1.42e-02} &  -0.062 & {\bf 3.60e-02} \\
$\lmstarnu$ & $\lamre$ & -0.183 & 3.99e-10 &  -0.017 & {\bf 5.54e-01} \\
$\log(sSFR)$ & $\lsfnu$ & -0.419 & 2.77e-50 &  -0.400 & 1.18e-45 \\
$\log(sSFR)$ & $\lamre$ &  0.543 & 2.05e-89 &   0.508 & 1.19e-76 \\
$\lsfnu$ & $\lamre$ & -0.163 & 2.33e-08 &   0.082 & 5.06e-03 \\
\hline
\end{tabular}
\end{table}

\begin{table}
	\centering
	\caption{The same as Table \ref{tab:corr_four_tests}, but for the GAMA regions only.}
	\label{tab:corr_four_restricted_gama}
	\begin{tabular}{ccrrrr} 
\hline
& & \multicolumn{2}{c}{full correlation} &\multicolumn{2}{c}{partial correlation}\\
A & B & $r$ & $p$-value & $r$ & $p$-value \\
\hline
$\lmstarnu$ & $\agel$ &  0.508 & 2.03e-48 &   0.476 & 1.10e-41 \\
$\lmstarnu$ & $\lsfnu$ &  0.075 & {\bf 4.47e-02} &  -0.060 & {\bf 1.12e-01} \\
$\lmstarnu$ & $\lamre$ & -0.207 & 2.24e-08 &   0.050 & {\bf 1.83e-01} \\
$\agel$ & $\lsfnu$ &  0.240 & 7.74e-11 &   0.243 & 4.41e-11 \\
$\agel$ & $\lamre$ & -0.475 & 1.15e-41 &  -0.443 & 1.01e-35 \\
$\lsfnu$ & $\lamre$ & -0.052 & {\bf 1.62e-01} &   0.075 & {\bf 4.53e-02} \\
\hline
$\lmstarnu$ & $\agem$ &  0.479 & 1.53e-42 &   0.444 & 6.66e-36 \\
$\lmstarnu$ & $\lsfnu$ &  0.075 & {\bf 4.47e-02} &  -0.001 & {\bf 9.69e-01} \\
$\lmstarnu$ & $\lamre$ & -0.207 & 2.24e-08 &  -0.078 & {\bf 3.60e-02} \\
$\agem$ & $\lsfnu$ &  0.158 & 2.08e-05 &   0.135 & 3.01e-04 \\
$\agem$ & $\lamre$ & -0.294 & 8.20e-16 &  -0.224 & 1.27e-09 \\
$\lsfnu$ & $\lamre$ & -0.052 & {\bf 1.62e-01} &  -0.006 & {\bf 8.70e-01} \\
\hline
$\lmstarnu$ & $\log(sSFR)$ & -0.413 & 1.41e-31 &  -0.354 & 4.17e-23 \\
$\lmstarnu$ & $\lsfnu$ &  0.088 & {\bf 1.67e-02} &  -0.017 & {\bf 6.52e-01} \\
$\lmstarnu$ & $\lamre$ & -0.214 & 4.71e-09 &   0.004 & {\bf 9.20e-01} \\
$\log(sSFR)$ & $\lsfnu$ & -0.249 & 7.39e-12 &  -0.243 & 2.44e-11 \\
$\log(sSFR)$ & $\lamre$ &  0.523 & 8.22e-53 &   0.492 & 6.26e-46 \\
$\lsfnu$ & $\lamre$ & -0.063 & {\bf 8.63e-02} &   0.081 & {\bf 2.81e-02} \\
\hline
\end{tabular}
\end{table}

\begin{table}
	\centering
	\caption{The same as Table \ref{tab:corr_four_tests}, but for the cluster regions only.}
	\label{tab:corr_four_restricted_cluster}
	\begin{tabular}{ccrrrr} 
\hline
& & \multicolumn{2}{c}{full correlation} &\multicolumn{2}{c}{partial correlation}\\
A & B & $r$ & $p$-value & $r$ & $p$-value \\
\hline
$\lmstarnu$ & $\agel$  &  0.343 & 4.28e-13 &   0.320 & 2.06e-11 \\
$\lmstarnu$ & $\lsfnu$ &  0.172 & 3.83e-04 &   0.101 & {\bf 3.96e-02} \\
$\lmstarnu$ & $\lamre$ & -0.137 & 4.99e-03 &   0.120 & {\bf 1.43e-02} \\
$\agel$  & $\lsfnu$ &  0.272 & 1.48e-08 &   0.109 & {\bf 2.61e-02} \\
$\agel$  & $\lamre$ & -0.629 & 9.39e-48 &  -0.604 & 4.68e-43 \\
$\lsfnu$ & $\lamre$ & -0.253 & 1.38e-07 &  -0.121 & {\bf 1.33e-02} \\
\hline
$\lmstarnu$ & $\agem$  &  0.405 & 4.91e-18 &   0.380 & 8.41e-16 \\
$\lmstarnu$ & $\lsfnu$ &  0.172 & 3.83e-04 &   0.103 & {\bf 3.53e-02} \\
$\lmstarnu$ & $\lamre$ & -0.137 & 4.99e-03 &   0.091 & \textbf{6.32e-02} \\
$\agem$  & $\lsfnu$ &  0.230 & 1.76e-06 &   0.078 & \textbf{1.10e-01} \\
$\agem$  & $\lamre$ & -0.485 & 3.61e-26 &  -0.452 & 1.91e-22 \\
$\lsfnu$ & $\lamre$ & -0.253 & 1.38e-07 &  -0.174 & 3.42e-04 \\
\hline
$\lmstarnu$ & $\log(sSFR)$ & -0.216 & 8.30e-06 &  -0.146 & 2.71e-03 \\
$\lmstarnu$ & $\lsfnu$ &  0.174 & 3.29e-04 &   0.118 & \textbf{1.56e-02} \\
$\lmstarnu$ & $\lamre$ & -0.137 & 4.93e-03 &  -0.002 & \textbf{9.66e-01} \\
$\log(sSFR)$ & $\lsfnu$ & -0.293 & 9.53e-10 &  -0.161 & 9.21e-04 \\
$\log(sSFR)$ & $\lamre$ &  0.577 & 1.22e-38 &   0.537 & 1.56e-32 \\
$\lsfnu$ & $\lamre$ & -0.259 & 7.66e-08 &  -0.114 & \textbf{2.00e-02} \\
\hline
\end{tabular}
\end{table}


\bsp	
\label{lastpage}
\end{document}